\documentclass[conference]{IEEEtran}
\IEEEoverridecommandlockouts

\def\BibTeX{{\rm B\kern-.05em{\sc i\kern-.025em b}\kern-.08emT\kern-.1667em\lower.7ex\hbox{E}\kern-.125emX}}

\usepackage{graphicx}
\usepackage{booktabs}
\usepackage{amsmath}
\usepackage{amsthm}
\usepackage{amssymb}
\usepackage{amsfonts}
\usepackage{epsfig}
\usepackage{graphicx}
\usepackage{subfigure}
\usepackage[bottom]{footmisc}
\usepackage{balance}
\usepackage{algpseudocode}
\usepackage{algorithmicx}
\usepackage{algorithm}
\usepackage{multirow}
\usepackage{xspace}
\usepackage[singlelinecheck=off]{caption}
\DeclareCaptionType{copyrightbox}
\usepackage{pifont}
\newtheorem{theor}{Theorem}

\newtheorem{problem}{Problem}

\newcommand{\spara}[1]{\smallskip\noindent{\bf #1}}

\newcommand{\squishlist}{
 \begin{list}{$\bullet$}
  {  \setlength{\itemsep}{0pt}
     \setlength{\parsep}{3pt}
     \setlength{\topsep}{3pt}
     \setlength{\partopsep}{0pt}
     \setlength{\leftmargin}{2em}
     \setlength{\labelwidth}{1.5em}
     \setlength{\labelsep}{0.5em}
} }
\newcommand{\squishlisttight}{
 \begin{list}{$\bullet$}
  { \setlength{\itemsep}{0pt}
    \setlength{\parsep}{0pt}
    \setlength{\topsep}{0pt}
    \setlength{\partopsep}{0pt}
    \setlength{\leftmargin}{2em}
    \setlength{\labelwidth}{1.5em}
    \setlength{\labelsep}{0.5em}
} }

\newcommand{\squishdesc}{
 \begin{list}{}
  {  \setlength{\itemsep}{0pt}
     \setlength{\parsep}{3pt}
     \setlength{\topsep}{3pt}
     \setlength{\partopsep}{0pt}
     \setlength{\leftmargin}{1em}
     \setlength{\labelwidth}{1.5em}
     \setlength{\labelsep}{0.5em}
} }

\newcommand{\squishend}{
  \end{list}
}

\newcommand{\eat}[1]{}

\newcommand{\NP}{\ensuremath{\mathbf{NP}}\xspace}
\newcommand{\sharpP}{\ensuremath{\mathbf{\#P}}\xspace}

\newcounter{ccc}

\newcommand{\bigO}{\mathcal{O}}

\usepackage{varwidth}
\usepackage{booktabs} 
\usepackage{bm}

\begin{document}

\title{\Large Maximizing Contrasting Opinions in Signed Social Networks}
	
\author{\IEEEauthorblockN{Kaivalya Rawal}
\IEEEauthorblockA{\textit{BITS-Pilani, India}\\
}
\and
\IEEEauthorblockN{Arijit Khan}
\IEEEauthorblockA{\textit{Nanyang Technological University, Singapore}\\
}
}

\maketitle

\begin{abstract}
The classic influence maximization problem finds a limited number of influential seed users
in a social network such that the expected number of influenced users in the network,
following an influence cascade model, is maximized. The problem has been
studied in different settings, with further generalization of the graph structure,
e.g., edge weights and polarities, target user categories, etc.
In this paper, we introduce a unique influence diffusion scenario involving a
population that split into two distinct groups, with opposing views. We aim at finding
the top-$k$ influential seed nodes so to simultaneously
maximize the adoption of two distinct,
antithetical opinions in the two groups, respectively.
Efficiently finding such influential users is essential in
a wide range of applications such as increasing
voter engagement and turnout, steering public debates and discussions
on societal issues with contentious opinions. We formulate this novel problem with the voter
model to simulate opinion diffusion and dynamics, and then design a linear-time
and exact algorithm {\sf COSiNeMax},
while also investigating the long-term opinion characteristics in the
network. Our experiments with several real-world datasets demonstrate the
effectiveness and efficiency of the proposed algorithm, compared to various baselines.
\end{abstract}

%



\vspace{-2mm}
\section{Introduction}
\label{introduction}
A central characteristic of social networks is that it facilitates rapid dissemination
of information among large groups of individuals \cite{CLC13,David2010}. Online social
networks, such as {\em Facebook}, {\em Twitter},
{\em LinkedIn}, {\em Flickr}, and {\em Digg} are used for spreading ideas and messages.
Users' behaviors and opinions are highly affected by their friends in social networks,
which is defined as the {\em social influence}.
Motivated by various real-world applications, e.g., viral marketing \cite{DR01}, social and political campaigning \cite{CKW15},
social influence studies have attracted extensive research attention. The classic influence
maximization problem \cite{KKT03,DR01} identifies the top-$k$ seed users in a social network such that the expected number
of influenced users in the network, starting from those seeds and following an influence diffusion model,
is maximized. The budget $k$ on the seed set size usually depends on how
many initial users the campaigner can directly influence by advertisements,
re-tweets from ``bots'', free samples and discounted prices.

In reality, societies are complex systems, and polarize into groups of individuals
with dramatically opposite perspectives. This phenomenon is also evident in online
social networks based on political affiliations, religious views, controversial
topics, personal biases and preferences \cite{GarimellaW17}. Therefore, each campaign is generally launched
and promoted with certain target audience in mind, e.g., all Republican
voters, people who prefer jazz over metal music, or Android over iPhones, etc.
Often, online campaigns have limited budgets and cannot afford to directly reach to all
members of their target population. In such scenarios, it is desirable to minimize
the number of seed users as permitted by the budget, while still maximizing the spread of the campaign
in the target audience.

Furthermore, due to the existence of subgroups with differing views, relationships between
social network users also include negative ones, such as foe, spite, and distrust relations.
Indeed, {\em signed social networks} containing both positive and negative relationships are ubiquitous \cite{Tang16}. For example, in the explicit category, users can directly tag the polarity (positive or negative)
to the relation between two users, e.g., {\em Epinions}, {\em Slashdot}, {\em Ebay}, and other online review and
news forums. In the implicit category, the relationship polarities can be mined from the interaction data between users, such as, in {\em Twitter}
a user $u$ may support some users whom she follows (positive) and be against the others (negative).
Following common sense and past literature on signed networks (including the {\em structural balance theory}) \cite{cartwright56,
LXCGSL14,VoterModelSigned,David2010}, we assume that positive relations
carry the influence in a positive manner, that is, a user would more likely trust and adopt her friends'
opinions. On the other hand, negative relations tend to carry influence in a reverse direction, i.e.,
if a user's foe chooses some opinion, the user would more likely be influenced to select the opposite one.
Our assumption supports the principles that {\em ``the friend of a friend is a friend''}, {\em ``the enemy of a friend is an enemy''},
{\em ``the friend of an enemy is an enemy''}, and {\em ``the enemy of an enemy is a friend''}. Ignoring such relationship
polarities between users and treating signed social networks as unsigned ones would result in over-estimation of positive
influence spread, thereby leading to lower-quality solutions. {\em Social influence can be further complicated when competing
campaigns are simultaneously spread over a signed social network. Therefore, influence and opinion dynamics in a
signed social network is a critical problem that, unfortunately, remains pretty much open}.

In this work, we investigate a {\em novel} influence diffusion problem: {\sf COSiNe} (\underline{C}ontrasting \underline{O}pinions
Maximization in a \underline{Si}gned Social \underline{Ne}twork). We aim to find a limited number of influential seed nodes which maximize the adoption of two distinct, antithetical opinions in two non-overlapping user groups with opposing views, respectively. The main objective behind such influence maximization is to create general awareness in a population by improving the quality of the debate on naturally contentious issues without inadvertently introducing prejudiced ideas.

\vspace{-1mm}
\spara{$\bullet$ Applications.} An ideal application of our problem would be to increase awareness about infrequently discussed issues that
are nonetheless controversial (such as capital punishment, nuclear energy, or affirmative action) --- in a target population that naturally
splits into two distinct ideological groups (such as democrats and republicans); in a forum that extensively debates topics and proposes mutually
agreeable solutions based on compromise, diversity, and inclusion (such as the United States Senate or House of Representatives).
Contrary to initial expectations, polarization of opinions and increased conflict can often be beneficial \cite{C19,S50,NatureWiki,Fortune,GovOpp, Columbia,Queens},
as discussed in the following.

The benefit of the conflicting opinions of various individuals collaborating together can be measured clearly on the online encyclopedia:
Wikipedia. Wikipedia uses a six-category scale (ranging from  ``stub'' to ``featured article'') to determine the quality of its articles, which are entirely
crowd-sourced. Controversial articles such as those on the Syrian Civil War, Israel/Palestine, or George W. Bush attract a higher number of edits. The
community debate can be seen on the ``talk page'' of each article. It has been found that higher polarization in the contributing community
is associated with higher article quality for a broad range of articles -- ranging from politics to science and social issues \cite{NatureWiki,C19}.

Increased diversity is often correlated with greater business performance \cite{HBR}. Similarly, disagreements amongst co-workers have been
found to improve the decision making capabilities at the organisation level; with a recent study from Columbia Business School stating ``cognitive conflict
(that is, differences in information, knowledge, and opinions) can be a critical source of competitive advantage'' \cite{Columbia}. Thus, there is a clear
merit in allowing and even encouraging different opinions about the same topic to flourish in a business setting. This can be leveraged to improve the
productivity of the organisation \cite{Fortune,Queens}. When dealt with correctly, such differences in thought and opinions are a force for good.

Lastly, we illustrate an example from the world of politics that is most similar to our ``ideal'' application scenario.
Unlike the American presidential system, in countries based upon the Westminster parliamentary system, there is an appointed head of government,
different from the head of the state, and an appointed head of opposition. This balance between the government and the opposition is considered
integral to the success of a functioning democracy in diverse countries such as in Britain and in India \cite{GovOpp}. An equivalent analysis
was made for the political system in the United States of America in 1950 by the American Political Science Association \cite{S50} which recommended
a stronger two party system in order to strengthen the democratic process. Both these analyses point to the importance of opposition in political discourse,
and go on to show that policies being enacted and implemented benefit from engagement, and even opposition. Meaningful discourse and spirited debate
requires people who inherently hold opposing beliefs on a given issue, and thus maximizing opposing influences can be beneficial for a legislative
body from the point of view of the general population.

\vspace{-1mm}
\spara{$\bullet$ Challenges and contributions.}
Contrasting opinions maximization, as required in our problem setting, is a non-trivial one.
First, one must employ an influence cascade model that has properties different from those for commercial, {\em one-time product purchasing} based marketing strategies.
For example, people's opinions change over time; thus, activation based models, such as
independent cascade (IC) and linear threshold (LT) models \cite{KKT03} are less appropriate in political contexts.
Second, in reality a signed social network might not be perfectly balanced \cite{VoterModelSigned}, that is, there may not exist a partition
$V_1, V_2$ of the node set $V$, such that all edges with $V_1$ and $V_2$ are positive and all edges across $V_1$ and $V_2$ are negative.
Such a network does not follow the social balance theory, and adds more complexity to the social influence cascade.

In this work, we employ the {\em voter model} \cite{VoterModel,holley1975,VoterModelUnsigned,VoterModelSigned}
to characterize influence diffusion in the two population groups of a social network.
We define our model such that opposite influences, when applied on the same user, cancel each other, leading to a decay
in the influence strength on any given user. Our model does not mandate that a user's choice be frozen upon one-time
activation, explicitly allowing the user to switch opinions at later times. Moreover, voter model, being a
stochastic one (it has a random walk based interpretation, which will be introduced in Section~\ref{sec:problem_formulation}),
can deal with signed networks that are not perfectly balanced. We then define our novel {\sf COSiNe} problem
(contrasting opinions maximization), and design an efficient, exact solution.

The main contributions of this paper are as follows.

\begin{itemize}
\setlength\itemsep{0.01em}
\vspace{-0.7mm}
\item We study the novel problem ({\sf COSiNe}) of finding the top-$k$ seed nodes that maximize the adoption of two distinct, antithetical
opinions in two given non-overlapping sets of target users, respectively, in a signed social network. We adapt the voter model
to formulate our problem in \S\ref{sec:problem_formulation}.
\item We design a linear-time, exact solution ({\sf COSiNeMax}) for our problem.
We demonstrate the correctness and derive time complexity of our algorithm in \S\ref{sec:alg}.
\item We further characterize two different long-term opinion dynamics in a signed social network
under extreme scenarios, and investigate how our proposed method, {\sf COSiNeMax} finds the seed
nodes intelligently under such extreme situations (\S\ref{sec:alg_long}).
\item We conduct a thorough experimental evaluation with several real-world signed social networks to demonstrate the effectiveness and efficiency of our algorithm,
compared to various baseline methods (\S\ref{sec:experiments}).
\end{itemize}

\vspace{-3mm}
\section{Preliminaries}
\label{sec:problem_formulation}
We model a social network as a signed, directed graph with edge weights: $G=(V,E,{\bf A})$,
where $V$ is the set of nodes (users), $E\subseteq V\times V$ is the set of
directed edges (links, connections, follower/followee relations, etc.),
and ${\bf A}$ is the weighted adjacency matrix with $A_{ij}\ne0$ when the
edge $(i,j)\in E$, with $A_{ij}$ being the weight of the edge $(i,j)$.
The weight $A_{ij}$ represents the strength of $j$'s influence on $i$.
Moreover, as we consider a signed graph, the adjacency matrix ${\bf A}$ may contain negative entries.
A positive entry $A_{ij}$ indicates a positive relation, i.e., $i$ considers $j$ as a friend
or $i$ trusts $j$, whereas a negative entry $A_{ij}$ denotes a negative relation, that is,
$i$ considers $j$ as a foe, or $i$ distrusts $j$.
The absolute value $|A_{ij}|$ represents the strength of this positive
or negative relation --- the higher, the stronger. We further
denote by ${\bf A^+}$ and ${\bf A^-}$ the (unsigned) matrices with only positive and negative entries
of ${\bf A}$, respectively. Thus, ${\bf A} = {\bf A^+} - {\bf A^-}$.
\vspace{-3mm}
\subsection{Information Diffusion Model}
\vspace{-1.5mm}
The voter model was first introduced in \cite{holley1975,VoterModel} to
investigate territorial conflicts between two species and more abstractly, the properties
of infinite systems of stochastic processes. It was then studied for maximizing
influence in unsigned networks \cite{VoterModelUnsigned} and
over signed networks \cite{VoterModelSigned}. We update the model from prior attempts in
order to more naturally simulate the spread of {\em two contrasting ideas}, $O_1$ and $O_2$,
{\em simultaneously} in the same network.

We associate with each node a floating point value $C$ in the range $[-1,1]$, that probabilistically
determines the node's adopted idea $O_1$ or $O_2$. The diffusion happens at discrete time steps,
and the $C$ value at every node can change with each time step. The opinion or idea adopted by node $i$
at time step $t$ is represented by $C_t(i)$: $C_t(i)\rightarrow 1$ implies that the user is likely to
adopt the idea $O_1$ at time step $t$, whereas $C_t(i)\rightarrow -1$ denotes that the user is likely to
adopt the idea $O_2$ at time step $t$. In particular, the probability of node $i$ adopting idea $O_1$
at time $t$ is defined as $p(O_1) = \frac{1+C_t(i)}{2}$, and the probability of $i$ adopting idea $O_2$
at time $t$ is $p(O_2) = \frac{1-C_t(i)}{2}$. The two probabilities are defined so that they
always sum up to one. In our voter model, each node starts uninfluenced in the beginning, i.e., $C_t = 0$ at time $t=0$,
except those nodes being influenced as seed nodes for ideas $O_1$ or $O_2$ by the campaigner.
For seed nodes, $C_0 = 1$ and $C_0 = -1$, respectively.

At every time step $t$, each node $i \in V$ adopts the idea of its outgoing neighbour $j \in V$
with probability $p = \frac{| A_{ij} |}{\Sigma_l | A_{il}|}$ if $A_{ij} > 0$, and adopts the opposite
idea if $A_{ij} < 0$. Formally,
\vspace{-1.5mm}
\begin{equation}
\label{eq:voter_model}
\begin{aligned}
\displaystyle & C_t(i)  \\
\displaystyle & = \Sigma_{j \in V}\left(\frac{A^{+}_{ij}}{\Sigma_{l \in V}|A_{il}|} C_{t-1}(j)\right) - \Sigma_{j \in V}\left(\frac{A^{-}_{ij}}{\Sigma_{l \in V}|A_{il}|} C_{t-1}(j)\right) \\
\displaystyle & = \Sigma_{j \in V}\frac{A^{+}_{ij}-A^{-}_{ij}}{\Sigma_{l \in V}|A_{il}|} C_{t-1}(j)
\vspace{-3mm}
\end{aligned}
\end{equation}

There is also an alternative, {\em random walk} interpretation of this voter model \cite{VoterModelSigned}. In this interpretation,
we consider a walk across the graph that starts at an arbitrary node $u$. At each time step, from the
current node $i$, an outgoing edge $i \rightarrow j$ is chosen with probability $p = \frac{| A_{ij} |}{\Sigma_l | A_{il}|}$
for the random walk. This walk is deemed to terminate at time $t$ on some node $v$. Then, according to the voter model,
$C_{t}(u) = C_0(v)$ if the path $u \rightarrow \cdots \rightarrow v$ has an even number of negative edges (a {\em positive path}),
and $C_{t}(u) = -C_0(v)$ if the path has an odd number of negative edges (a {\em negative path}).

By defining the voter model this way, opposite influences on a particular node tend to ``cancel'' out.
The voter model also allows the opinion of a user to flip between two contrasting ideas, based on her
neighbors' influences. Thus, our voter model is different from one-time, activation-based influence propagation models
(e.g., independent cascade (IC) and linear threshold (LT) models \cite{KKT03}), and we employ it to study opinion
diffusion and formation in online signed social networks.
\vspace{-1.5mm}
\subsection{Problem Statement}
\vspace{-1mm}
Two non-overlapping groups $V_1$ and $V_2$ among the social network users are
given as an input to our problem, such that, $V_1 \cap V_2 =\phi$ and
$V_1 \cup V_2 \subseteq V$. The campaigner aims at influencing all nodes in $V_1$
with the idea $O_1$, and all nodes in $V_2$ with the idea $O_2$. Clearly, the users outside
both the groups $V_1$ and $V_2$ have no business value to the campaigner.

We define an {\em opinion vector} ${\bf C_t}$, according to the opinions of all the nodes in our network
at any specific time $t$. Thus, for a network with $|V|=n$ nodes:
\vspace{-3mm}
\begin{equation}
{\bf C_t} =
\left [
\begin{matrix}
C_t(0) \\ C_t(1) \\ \vdots \\ C_t(n-1)
\end{matrix}
\right ]
\end{equation}
\vspace{-2mm}

The voter model can be described in matrix form in terms of the opinion vector and a
{\em transition matrix} ${\bf P}={\bf D}^{-1}{\bf A}$. Here, ${\bf D}$ is a diagonal matrix that consists of all
entries of $({\bf A^+} + {\bf A^-}) \cdot {\bf 1}$ in its diagonal. From Equation~\ref{eq:voter_model},
we get:
\vspace{-2mm}
\begin{equation}
\label{eq:matrix_form}
\begin{aligned}
& C_t(i)  = \Sigma_{j \in V}\frac{A^{+}_{ij}-A^{-}_{ij}}{\Sigma_{l \in V}|A_{il}|} C_{t-1}(j) \\
\implies & C_t(i)  = \Sigma_{j \in V}\frac{A_{ij}}{\Sigma_{l \in V}|A_{il}|} C_{t-1}(j) \\
\implies & {\bf C_t}  = {\bf D}^{-1} {\bf A C_{t-1}} = {\bf P C_{t-1}} = {\bf P}^t {\bf C_{0}}
\end{aligned}
\end{equation}
\vspace{-3mm}

Similar to the opinion vector, we define a partition vector ${\bm \rho}$ to
describe two target populations $V_1$ and $V_2$. We define element $\rho_i$ in this vector,
for each node $i\in V$, as below:
\vspace{-2mm}
\begin{equation}
\rho_i =
\begin{cases}
 +1    \;\;\;   \dots  \;\;\;  \text{if}  \;\;  i \in V_1 \\
 -1    \;\;\;   \dots  \;\;\;  \text{if}  \;\;  i \in V_2 \\
 0 \;\;\;\; \dots \;\;\; \text{if} \;\; i \in V \land i \notin (V_1 \cup V_2)
\end{cases}
\end{equation}
\vspace{-2mm}

The effectiveness $\epsilon_t$ of the advertising campaign across both target
populations can now be measured by using the scalar product formula
$\epsilon_t = {\bm \rho^T} \cdot {\bf C_t}$. This promotes opinion $O_1$ in
partition $V_1$ and opinion $O_2$ in partition $V_2$, while also penalising
the reverse situation, that is, $O_1$ in $V_2$ and $O_2$ in $V_1$.
The formulation correctly ignores the opinions of the nodes that do not
belong in either $V_1$ or $V_2$, that the campaigner is agnostic towards.
It is worth noting that $\epsilon_t$ is a function of three parameters.
(1) Future time step $t$: input to the problem, (2) ${\bm \rho}$: which defines
two non-overlapping target groups and is provided as an input to the problem,
and (3) $C_0$: the seed set that needs to be determined.

We consider budget $k$ on the number of seed nodes, which is an input parameter.
We are now ready to define our problem.
\vspace{-3mm}
\begin{problem}
\label{ref:op_max}
[COSiNe] Given a signed, directed graph with edge weights: $G=(V,E,{\bf A})$, a future time step $t>0$,
${\bm \rho}$ vector which defines two non-overlapping target groups $V_1, V_2$ for two contrasting ideas $O_1$ and $O_2$,
respectively, and a budget $k$ on the total number of seed nodes, find the top-$k$ seed nodes, together
with their advertisement types (between $O_1$ and $O_2$), such that the effectiveness $\epsilon_t = {\bm \rho^T} \cdot {\bf C_t}$
of the campaign is maximized.
\end{problem}

\vspace{-3mm}
\section{Algorithm: Short-Term Opinions Maximization}
\label{sec:alg}
\vspace{-1.5mm}
In this section, we design an efficient and {\em exact} algorithm for
the {\sf COSiNe} problem and with a given, finite time step $t>0$.
We refer to this as ``short-term'' since $t$ could be small and we
do not look for characteristics of the opinion dynamics as $t\rightarrow\infty$.
The long-term case will be discussed in Section~\ref{sec:alg_long}.

Our strategy for finding the most influential seed nodes is as follows. We compute the amount of influence of each node
on the rest of the network at time $t$. It turns out that, according to our voter model, selecting the top-$k$ individually
most influential nodes as the seed nodes is equivalent to the set of $k$ nodes with the highest influence. The correctness
of our algorithm is proved in Section~\ref{sec:correct}.

Our complete algorithm, {\sf COSiNeMax} is given in Algorithm~\ref{alg:cosinemax}.
To find the individual influence power $\epsilon(i)$ of each node $i\in V$, we simulate random walks
in the reverse direction of the actual influence diffusion (Lines 1-14).
The number of walks terminating at a specific node can thus be used as a measure of the node's
ability to influence other nodes, based on our voter model. We next select the top-$k$ nodes
having the maximum absolute influence power individually as the seed set (Lines 15-37).
Furthermore, for a seed node $j$, if $\epsilon(j)$ is positive, it is influenced with
idea $O_1$; otherwise the seed node is influenced with $O_2$ (Lines 29-33).
\vspace{-4.3mm}
\subsection{Proof of Correctness}
\label{sec:correct}
\vspace{-2mm}
We prove the correctness of Algorithm~\ref{alg:cosinemax} in two steps. First, we show that
the aggregate of the individual influence of $k$ nodes is identical to the influence strength
of the set consisting of the same $k$ nodes together (Theorem~\ref{th:aggregate}). Second, we demonstrate that the seed set
formed by the top-$k$ nodes as selected by Algorithm~\ref{alg:cosinemax} is indeed the best seed
set given inputs $G$, $t$, $k$, and ${\bm \rho}$ (Theorem~\ref{th:opt}).
\vspace{-2.3mm}
\begin{theor}
Let $\epsilon_t = {\bm \rho^T} \cdot {\bf C_t}$ be the total influence of a seed set $\Omega$ consisting of
$k$ nodes. We denote by $\epsilon_t(i)$ the individual influence of a node $i\in \Omega$. Then,
$\epsilon_t = \sum_{i\in \Omega}\epsilon_t(i)$.
\label{th:aggregate}
\end{theor}
\vspace{-4mm}
\begin{proof} \renewcommand{\qedsymbol}{}
We denote by $\Omega$ the seed set with $k$ nodes. The subset of seed nodes influenced by the idea $O_1$ is denoted as
$\Omega^+$, whereas the subset of seed nodes influenced by the idea $O_2$ is denoted as $\Omega^-$.
Clearly, $\Omega_1 \cap \Omega_2 =\phi$ and $\Omega_1 \cup \Omega_2 = \Omega$. Let $\epsilon_t$ be the
total influence by the seed set $\Omega$, whereas we represent by $\epsilon_i(t)$ the individual influence
when the seed set consists of the single node $i \in \Omega$.

Consider three vectors ${\bm e_1}$, ${\bm e_2}$, and ${\bm e_i}$, each having dimensionality $|V|$.
They represent various subsets of $\Omega$: ${\bm e_i}$ consists of $|V|-1$ zeros, with only the $i$-th element
being $\pm 1$ (depending on whether $i$ has been influenced with idea $O_1$ or $O_2$, respectively),
representing the singleton set $\{i\}$.
Analogously, ${\bm e_1}$ consists of $+1$ corresponding to all nodes in the set $\Omega_1$,
and ${\bm e_2}$ consists of $-1$ for all nodes in the set $\Omega_2$. The rest of the elements
in ${\bm e_1}$ and ${\bm e_2}$ are zeros. Formally,
\vspace{-2mm}
\begin{equation}
\begin{aligned}
e_1(j) & =
\begin{cases}
  0 \;\;\;\quad \text{if}\; j \notin \Omega_1 \\
  +1 \;\quad     \text{if}\; j \in \Omega_1
\end{cases}
e_2(j)  =
\begin{cases}
  0 \;\;\;\quad \text{if}\; j \notin \Omega_2 \\
  -1 \;\quad    \text{if}\; j \in \Omega_2 \\
\end{cases}\\
e_i(j) & =
\begin{cases}
  0 \;\;\;\quad \text{if}\; j \neq i \\
  +1 \;\quad    \text{if}\; j = i,\;j \in \Omega_1 \\
  -1 \;\quad    \text{if}\; j = i,\;j \in \Omega_2
  \vspace{-1mm}
\end{cases}
\vspace{-4mm}
\end{aligned}
\end{equation}
Thus, ${\bm e} = {\bm e_1} + {\bm e_2}$ is the vector denoting the seed set $\Omega = \Omega_1\cup\Omega_2$.
Next, we derive the following.
\vspace{-6mm}
\begin{equation}
\label{eq:individual}
\begin{aligned}
  \epsilon &= {\bm \rho}^T \cdot {\bm C_t} ={\bm \rho}^T \cdot ({\bm P^t e})  \qquad \vartriangleright\text{Following Equation~\ref{eq:matrix_form}}\\
     &={\bm \rho}^T \cdot {\bm P^t} ({\bm e_1} + {\bm e_2}) ={\bm \rho}^T \cdot {\bm P^t} \left(\Sigma_{i \in \Omega_1} \left({\bm e_i}\right) + \Sigma_{i \in \Omega_2} \left({\bm e_i}\right)\right)\\
     &=\Sigma_{i \in \Omega} ({\bm \rho}^T {\bm P^t e_i}) = \Sigma_{i \in \Omega} ({\bm \rho}^T {\bm C_t(i)} ) \qquad \vartriangleright\text{Following Equation~\ref{eq:matrix_form}} \\
     &= \Sigma_{i \in \Omega} \epsilon_i
     \vspace{-3.5mm}
\end{aligned}
\vspace{-2mm}
\end{equation}
Hence, the theorem.
\end{proof}
\begin{algorithm}[tb!]
\caption{{\sf COSiNeMax}: Maximize Contrasting Opinions}\label{alg:cosinemax}
\begin{algorithmic}[1]
\Require  Signed graph $G=(V,E,{\bf A})$; time step $t>0$; ${\bm \rho}$ vector to define
two non-overlapping target groups $V_1, V_2$ for two contrasting ideas $O_1$, $O_2$, respectively; budget $k$
\Ensure Set $\Omega$ of top-$k$ nodes, with their advertisement types (between $O_1$ and $O_2$), that maximizes $\epsilon_t = {\bm \rho^T} \cdot {\bf C_t}$
\State ${\bf P}={\bf D}^{-1}{\bf A}$ \Comment{Transition Matrix of $G$}
\State ${\bm \epsilon} \gets [0, 0, 0, \ldots, 0]$ \Comment{Initialise row vector of size $|V|$}
\For{$i\gets 1, |V|$}
  \If{$i \in V_1$}
	\State $\epsilon[i] \gets +1$
  \ElsIf{$i \in V_2$}
     \State $\epsilon[i] \gets -1$
  \Else
    \State $\epsilon[i] \gets 0$
 \EndIf
\EndFor
\For{$i\gets 1, t$}
\State ${\bm \epsilon} = {\bm \epsilon} \cdot {\bf P}$
\EndFor	
\newline $\qquad$ \Comment{${\bm\epsilon}$ is distribution of reverse random walks at time $t$}
\State $\Omega \gets \Phi$ \Comment{$\Omega$ is a set of tuples $\langle i \in V, \tau(i)\rangle$}
\newline $\qquad \qquad$ \Comment{$\tau(i)$ denotes the individual influence of node $i$}
\For{$j \gets 1, |V|$}
  \If{$size(\Omega) \leq k$}
	\State insert $(\Omega, \langle j, |\epsilon[j]|\rangle)$
    \If{$\epsilon(j)>0$}
      \State $Opinion(j)\gets O_1$
    \Else
      \State $Opinion(j)\gets O_2$
    \EndIf
  \Else
    \State $\langle i, \tau(i)\rangle \gets \min(\Omega)$ \Comment{$\min$ is based on $\tau()$ values}
    \If{$|\epsilon[j]| > \tau(i)$}
      \State remove $(\Omega, \langle i, \tau(i)\rangle)$
	  \State insert $(\Omega, \langle j, |\epsilon[j]|\rangle)$
      \If{$\epsilon(j)>0$}
        \State $Opinion(j)\gets O_1$
      \Else
        \State $Opinion(j)\gets O_2$
      \EndIf
    \EndIf
  \EndIf
\EndFor
\State \textbf{return} $\Omega, Opinion(i:i\in\Omega)$ \Comment{Optimal seed nodes, with their advertisement types between $O_1$ and $O_2$}
\end{algorithmic}
\end{algorithm}
\vspace{-4.5mm}
\begin{theor}
The seed set $\Omega$, consisting of the top-$k$ individually most influential nodes as selected by Algorithm~\ref{alg:cosinemax},
is the optimal seed set having size $k$.
\label{th:opt}
\end{theor}
\vspace{-4.5mm}
\begin{proof} \renewcommand{\qedsymbol}{}
Notice that Algorithm~\ref{alg:cosinemax} selects the top-$k$ individually most influential nodes into the seed set $\Omega$.
Therefore, the following holds: $\epsilon_j \geq \epsilon_i$ for all nodes $i,j\in V$, such $j\in \Omega$ and $i \not\in \Omega$.

We demonstrate that for any other seed set $\Omega'$, such that $\Omega' \neq \Omega$, $|\Omega'| = |\Omega|$ cannot have more influence
than that of $\Omega$. Let us define $\omega' = \Omega' \setminus \Omega$, $\omega = \Omega \setminus \Omega'$,
and $o = \Omega' \cap \Omega$. Note that since the size of both $\Omega$ and $\Omega'$ is $k$, $|\omega'| = |\omega|$.

We prove by contradiction: Following Theorem~\ref{th:aggregate}, and if possible, we assume that $\Sigma_{i \in \Omega'} \epsilon_i > \Sigma_{j \in \Omega} \epsilon_j$.
Then, we get:
\vspace{-2mm}
\begin{equation}
\begin{aligned}
& \Sigma_{i \in \Omega'} \epsilon_i  > \Sigma_{j \in \Omega} \epsilon_j \\
\implies &\Sigma_{i \in \omega' \cup o} \epsilon_i  > \Sigma_{j \in \omega \cup o} \epsilon_j \\
\implies & \Sigma_{i \in \omega'} \epsilon_i + \Sigma_{i \in o} \epsilon_i  > \Sigma_{j \in \omega} \epsilon_j + \Sigma_{j \in o} \epsilon_j \\
\implies & \Sigma_{i \in \omega'} \epsilon_i  > \Sigma_{j \in \omega} \epsilon_j \\
\implies & \exists(i \in \omega', j \in \omega) \quad \text{such that} \quad \epsilon_i  > \epsilon_j \\
\implies & \exists(i \notin \Omega, j \in \Omega) \quad \text{such that} \quad \epsilon_i  > \epsilon_j
\end{aligned}
\end{equation}
This contradicts that Algorithm~\ref{alg:cosinemax} selects the top-$k$ individually most influential nodes into the seed set $\Omega$.
Hence, the theorem.
\end{proof}
\vspace{-7mm}
\subsection{Time Complexity Analysis}
\label{sec:complexity}
\vspace{-1.5mm}
Time complexity of our algorithm is: $\bigO(|E|t)$ as follows.

\vspace{-1mm}
\spara{Transition matrix calculation.}
Line 1 finds the transition matrix ${\bf P}$. This is an $\bigO(|E|)$ operation,
as it involves using the element-wise absolute values in ${\bf A}$, calculating ${\bf D}$, and finally computing ${\bf D}^{-1} \cdot {\bf A}$.
Note that real-world networks are generally sparse, thus ${\bf A}$ can be represented as a sparse matrix with $|E|$ non-zero elements.
Inverting ${\bf D}$ is an $\bigO(|V|)$ operation, since ${\bf D}$ is a diagonal matrix:
The inverse of a diagonal matrix is obtained by replacing each element in the diagonal with its reciprocal.
Finally, ${\bf D}^{-1} \cdot {\bf A}$ can be computed in $\bigO(|E|)$ time via sparse matrix multiplication, as each diagonal element of ${\bf D}^{-1}$ is
multiplied with exactly one element of ${\bf A}$, and this forms a non-zero element in the transition matrix ${\bf P}$.
Moreover, it is easy to verify that ${\bf P}$ will have $|E|$ non-zero elements.

\vspace{-1mm}
\spara{Initialisation of $\epsilon$.}
This requires time $\bigO(|V|)$ in lines 3-11.

\vspace{-1mm}
\spara{Random walk simulation.}
The slowest step in the algorithm is random walk simulation in lines 12-14.
In this phase, we require $\bigO(|E|t)$ time. Since $\epsilon$ is a
one dimensional vector, each multiplication in line 13 costs $\bigO(|E|)$
due to sparse matrix multiplication, and this operation is repeated $t$ times.

\vspace{-1mm}
\spara{Seed set selection.}
Finally, in lines 15-37 we select the top-$k$ nodes with the individually highest
absolute influence power. This is similar to choosing the top-$k$ elements in an
unordered list, and can be accomplished in $\bigO(|V|\log k)$ time.

Thus, time complexity of our algorithm is bounded by the random walk simulation,
and the time complexity is: $\bigO(|E|t)$, which is linear in the size of the input graph.

\vspace{-2mm}
\section{Long-Term Opinions Formulation}
\label{sec:alg_long}
\vspace{-1mm}
We now turn our attention to the long-term scenario, that is, opinion dynamics as $t\rightarrow\infty$.
In particular, we consider two extreme scenarios with respect to the two non-overlapping groups
$V_1$ and $V_2$ in the signed social network. For simplicity, in this section we shall assume
that $V_1 \cup V_2 = V$ and the graph is strongly connected.

\noindent {\bf $\bullet$ Socially balanced partitions:} With respect to partitions $V_1, V_2$, all intra-partition edges are positive, and all inter-partition edges are negative.
{\bf $\bullet$ Socially anti-balanced partitions:} With respect to partitions $V_1, V_2$, all intra-partition edges are negative, and all inter-partition edges are positive.

\spara{Remarks.} First, even though most real-world datasets do not exactly fall under the above two categories, a real-world
network could resemble one of them. For example, we observe that the {\em Tagged} dataset \cite{Tagged} that we use in our experiments, has more than three times
as many positive inter-partition edges than all other kinds of edges combined, thereby making these partitions close to socially anti-balanced partitions. By analyzing
the long-term opinion dynamics for the two categories, we demonstrate how intelligently our algorithm finds the seed nodes even under such extreme situations.
Second, we employ our algorithm, {\sf COSiNeMax} in all scenarios, as its optimality has been proved in \S\ref{sec:correct} irrespective of future time step $t$ (i.e.,
short-term vs. long-term), graph structures, and node partitions.

For ease of discussion, we define a {\em signed path} in a signed, directed social network as a sequence of nodes with the edges being directed from each node to the following one.
The {\em length of the path} is the total number of directed edges in it. The {\em sign of a path} is positive if there is an even number of negative edges along the path;
otherwise the sign of a path is negative.
\vspace{-4mm}
\subsection{Socially Balanced Partitions}
Recall that the campaigner's objective is as follows: At time step $t$, all
nodes in $V_1$ will adopt opinion $O_1$, and nodes in $V_2$ will adopt opinion $O_2$.
We next show that {\em if the input partitions are socially balanced, then by following our algorithm,
at  $t\rightarrow\infty$, indeed nodes in $V_1$ will adopt opinion $O_1$ and nodes in $V_2$ will adopt $O_2$}.

To prove this, it is easy to verify that all paths that begin and end in the same partition have positive signs (due to even number of
negative edges on those paths). Analogously, all paths that begin in one partition and end in the other partition must have negative signs
because of odd number of negative edges on them. This has two implications as given below.

First, {\em {\sf COSiNeMax} will select all seed nodes of $O_1$ only from the users in $V_1$, and all seeds for $O_2$ only from $V_2$}.
This is because in Lines 4-7 of Algorithm~\ref{alg:cosinemax}, all nodes in $V_1$ starts as positive,
and in partition $V_2$ all nodes starts as negative (at $t=0$). Now, repeated multiplications with
the transition matrix ${\bf P}$ (Lines 12-14) can be considered as a union of random walks. Therefore,
at any arbitrary future time step $t$, all nodes in $V_1$ would remain positive, because
all random walks starting at $V_1$ and also ending at $V_1$ must consist of only positive paths.
Similarly, at any arbitrary future time step $t$, all nodes in $V_2$ would remain negative.
Now, in Lines 29-33, the seed nodes are influenced based on their final sign, that is, if positive
then influenced with opinion $O_1$, and otherwise with opinion $O_2$. This concludes that the
seed nodes for $O_1$ will only be selected from group $V_1$, and those for $O_2$ will be picked only from $V_2$.

Second, {\em for socially balanced partitions, if all seeds of $O_1$ are from $V_1$, and all seeds for $O_2$ are from $V_2$,
then at $t\rightarrow\infty$, nodes in $V_1$ will adopt opinion $O_1$ and nodes in $V_2$ will adopt $O_2$}. This holds
because each path from any seed in $V_1$ to some other node in $V_1$ will always be a positive path, thereby
carrying the same opinion as that of the seed (i.e., $O_1$), whereas every path from a seed in $V_2$
to some other node in $V_1$ will be a negative path, thereby carrying the opposite opinion to that of the seed (i.e., also $O_1$).
\vspace{-3mm}
\subsection{Socially Anti-balanced Partitions}
\vspace{-1mm}
We show that {\em if all seeds of $O_1$ are from $V_1$, all seeds for $O_2$ are from $V_2$, and when $t \rightarrow \infty$, then anti-balanced partitions switch opinions between $O_1$ and $O_2$ at even and odd time steps, respectively}.
\vspace{-0.5mm}
\subsubsection{Even time steps}
For even time steps, we consider paths of even lengths. Among such paths, all paths that begin and end in the same partition have positive signs (due to even number
of negative edges), and all paths that begin and end in different partitions have negative signs (due to odd number of negative edges).
Hence, this is identical to the situation in socially balanced partitions, and similar results hold. In other words, {\bf (1)}
{\sf COSiNeMax} will select all seed nodes of $O_1$ only from the users in $V_1$, and all seeds for $O_2$ only from $V_2$.
{\bf (2)} For socially anti-balanced partitions, if all seeds of $O_1$ are from $V_1$, and all seeds for $O_2$ are from $V_2$,
then at $t\rightarrow\infty$, with $t$ being even, nodes in $V_1$ will adopt opinion $O_1$ and nodes in $V_2$ will adopt $O_2$.
\vspace{-0.5mm}
\subsubsection{Odd time steps}
For odd time steps (with $t\rightarrow\infty$), one can follow similar reasoning to show that the opposite case arises.
We now consider paths of odd lengths. Among such paths, all paths that end in the same partition as they began have negative signs (due to odd number of negative edges),
and all paths that end in the opposite partition as they began have positive signs (due to even number of negative edges).
This results in swapping of opinions for the two partitions, relative to the ones in an even time step.

{\em Notice that {\sf COSiNeMax} intelligently selects seed nodes: When the objective is to maximize the adoption of $O_1$ at $V_1$
and $O_2$ at $V_2$ in an odd time step, in anti-balanced partitions as $t\rightarrow\infty$, {\sf COSiNeMax} will select all seed nodes of $O_1$ only from the users in $V_2$,
and all seeds for $O_2$ only from $V_1$}. 
\vspace{-2mm}
\section{Experimental Results}
\label{sec:experiments}
\vspace{-2mm}
We show empirical results to demonstrate effectiveness and
efficiency of our solution, and compare it with three baselines. We analyze sensitivity of
{\sf COSiNeMax} by varying several parameters, e.g., number of seed and targets, time steps.
\vspace{-1mm}
\subsection{Environment Setup}
\label{sec:setup}
\vspace{-1mm}
Our code is implemented in Python, using sparse matrix operations from the {\em scipy} library,
and the experiments were performed on a single core of a 16GB, 1.8GHz, Intel i7-8550U processor.
Each experimental result is averaged over 10 runs. Our {\em source code} and {\em datasets} are publicly available at: github.com/COSiNe Max/COSiNe-Max
and drive.google.com/drive/folders/1hHn14eYehzRp8nk\_sup RfnhahXDDjjmn?usp=sharing, respectively.
\vspace{-7mm}
\subsubsection{Datasets}
\vspace{-2mm}
We summarize our datasets in Table~\ref{tab:data}.
\spara{(1) Epinions.} This social network dataset is extracted from the product review website epinions.com,
where users may trust or distrust others \cite{Epinions}. It is a signed and directed network:
A user trusting another is represented with an edge of weight $+1$, and distrusting another is denoted by weight $-1$.
The products being reviewed fall into one of 34 unique verticals, and we, uniformly at random,
partition these verticals into two categories. The nodes are then split into two non-overlapping partitions $V_1$ and $V_2$ depending
on the product categories that they review.
\spara{(2) GitHub.} The dataset (blog.github.com/2009-07-29-the-2009-github-contest) is
extracted from an anonymized dataset of user-repository
interactions on github.com, utilising information about users "watching" other's repositories.
We classify users into partitions $V_1, V_2$ based on whether the most used language in their watched
repositories is among the top-10 most popular languages following {\sf TIOBE} index:
{tiobe.com/tiobe-index/programming-langu ages-definition/}. We connect any two users
in the network with a bidirectional edge if they watch the same repository,
with edge weight inversely proportional to the number of watchers for that repository.
The sign of this edge is positive if both nodes view more single-language repositories (or, both view more multi-language repositories),
and negative otherwise (i.e., one views single-language repositories and the other views multi-language repositories).
The signed edge weight distribution is shown in Figure~\ref{fig:git_distribution}.
\spara{(3) Tagged.} Our largest real-life dataset is collected from the online social network tagged.com \cite{Tagged}.
The nodes are partitioned into $V_1$ and $V_2$ using anonymized gender metadata. Moreover, each edge of the network
belongs to one of seven categories. This categorical information is converted into a signed edge weight as given in Table~\ref{tab:tagged}:
The intuition is to have many modestly weighted positive and negative edges (i.e., edge weights between -0.5 to 0.8),
and only a few edges with very high positive and negative edge wights (i.e., edge weights -1.0 or +1.0).
\begin{table}[tb!]
\centering
\scriptsize
\vspace{-4mm}
\caption{\small Dataset characteristics}
\label{tab:data}
\vspace{-2mm}
\begin{tabular}{|l|r|r|r|r|}
\hline
\multicolumn{1}{|c|}{\textbf{Dataset}}& \multicolumn{1}{c|}{\textbf{\#Nodes}} & \multicolumn{1}{c|}{\textbf{\#Edges}} & \multicolumn{1}{c|}{\textbf{\#Positive Edges}} & \multicolumn{1}{c|}{\textbf{\#Negative Edges}} \\ \hline
{\em Epinions} & 132\,585 & 701\,926 & 605\,854 (86\%) & 96\,072 (14\%)  \\ \hline
{\em GitHub}   & 44\,914  & 44\,100\,700 & 26\,185\,530 (59\%) & 17\,915\,170 (41\%)   \\ \hline
{\em Tagged}   & 5\,607\,448 & 546\,799\,071 & 443\,895\,613 (81\%) & 102\,903\,458 (19\%) \\ \hline
\end{tabular}
\vspace{-2mm}
\end{table}
\begin{table}[tb!]
\begin{varwidth}[b]{0.45\linewidth}
\centering
\vspace{-1mm}
\caption{\small {\em Tagged}: Signed edge weight distribution}
\label{tab:tagged}
\vspace{-1mm}
\begin{scriptsize}
\begin{tabular}{|c|c|c|}
\hline
\textbf{Cat.}  & \multicolumn{1}{c|}{\textbf{Weight}} & \multicolumn{1}{c|}{\textbf{\#Edges}} \\
\hline
1 &-1.0 & 5\,762K (0.67\%)   \\
2 &-0.9 & 9\,361K (1.09\%)   \\
3 &-0.5 & 139\,379K (16.24\%)  \\
4 &-0.1 & 202\,003K (23.53\%)  \\
5 &0.3  & 150\,877K (17.58\%)  \\
6 &0.8  & 350\,724K (40.87\%)  \\
7 &1.0  & 137K (0.02\%) \\ \hline
\end{tabular}
\end{scriptsize}
\end{varwidth}%
\hfill
\begin{minipage}[b]{0.45\linewidth}
\centering
\vspace{-2mm}
\includegraphics[scale=0.38]{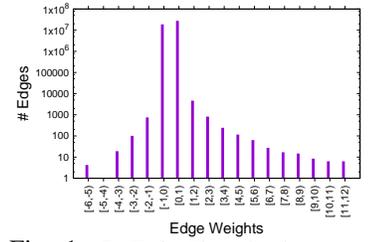}
\vspace{-6mm}
\captionof{figure}{\small {\em GitHub}: Signed edge weight distribution}
  \label{fig:git_distribution}
\vspace{-5mm}
\end{minipage}
\vspace{-3mm}
\end{table}
%
%
\subsubsection{Competing Methods}
\label{sec:compete}
We compare the proposed {\sf COSiNeMax} method (Algorithm~\ref{alg:cosinemax})
with three baselines.
{\bf (1) Random.} Uniformly at random selection of $k$ seed nodes.
{\bf (2) Degree.} The top-$k$ nodes with the highest out-degrees.
{\bf (3) Individual InfMax.} In this baseline approach,
we follow the {\em voter model} over signed networks \cite{VoterModelSigned}, however we consider each target set
separately. That is, we first compute the top-$\lfloor k/2 \rfloor$ seed nodes so to maximize the spread of the idea $O_1$ in the target partition $V_1$.
Next, we find another top-$\lfloor k/2 \rfloor$ seed nodes that maximize the spread of the idea $O_2$ within the target set $V_2$. Therefore, by comparing
with the {\em Individual Influence Maximization} approach as described above, we demonstrate the improvements due to our algorithm {\sf COSiNeMax}, which
returns the top-$k$ optimal seed nodes considering the spread of two contrasting ideas $O_1$ and $O_2$ simultaneously.

For each baseline, at $t=0$ we target a seed node $i$ with idea $O_1$ if $i \in V_1$, and with $O_2$ if $i \in V_2$.
\begin{figure*}[t!]
\centering
\subfigure [{\small {\em Epinions}, \#seeds=5\% of all users}]{
\includegraphics[scale=0.29]{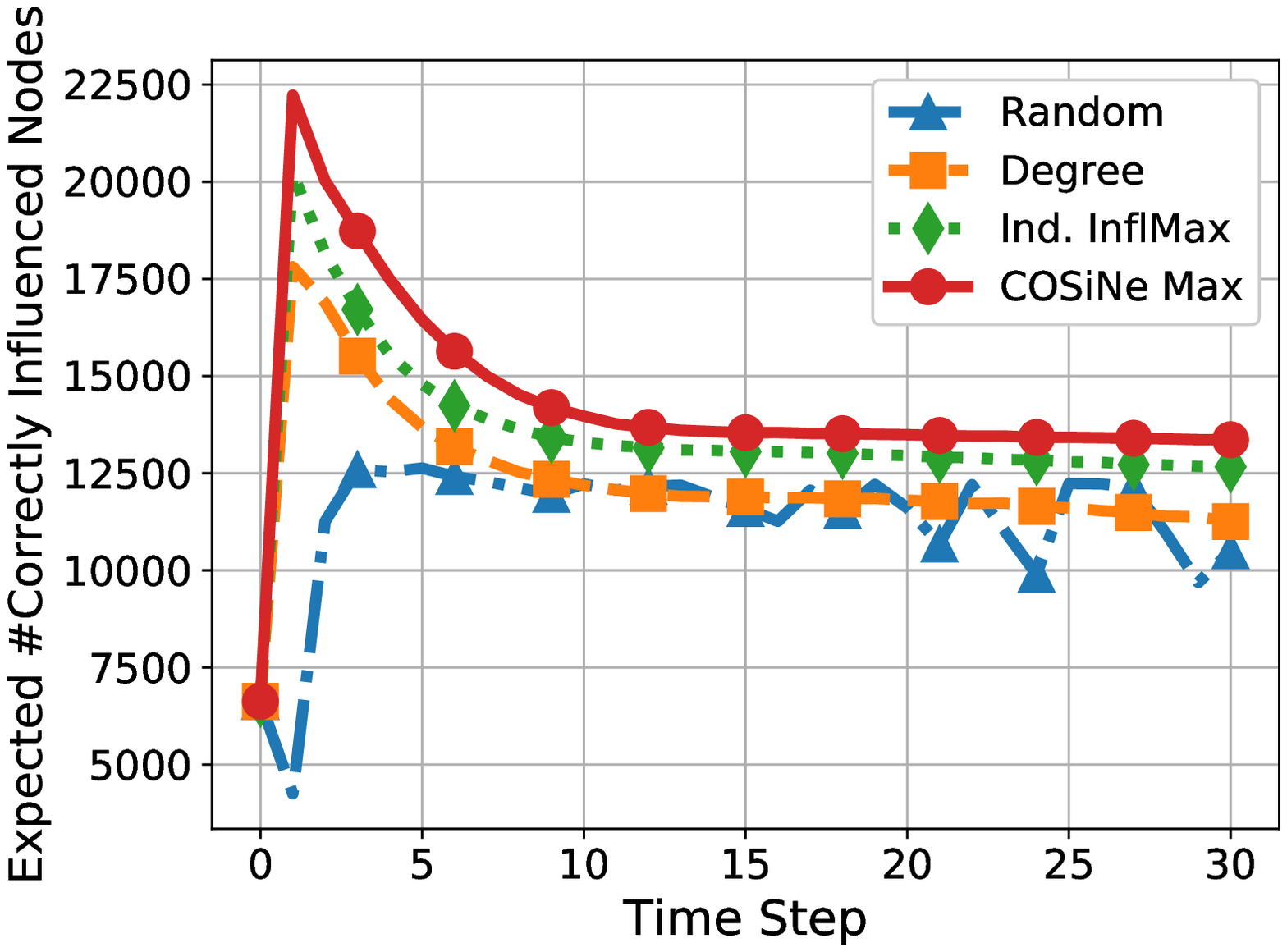}
\label{fig:boat_ep}
}
\subfigure [{\small {\em GitHub}, \#seeds=5\% of all users}] {
\includegraphics[scale=0.29]{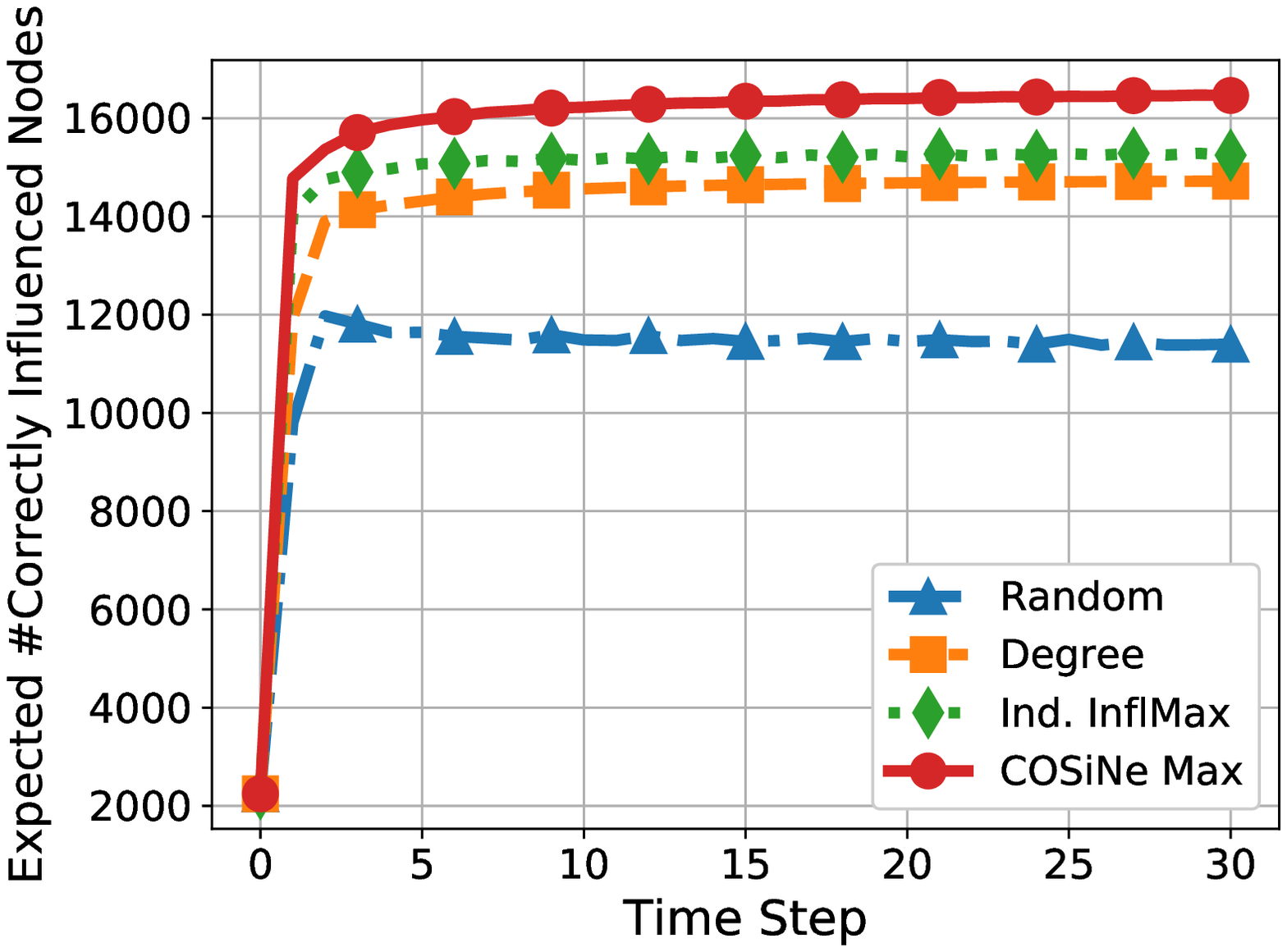}
\label{fig:boat_gh}
}
\subfigure [{\small {\em Tagged}, \#seeds=1\% of all users}] {
\includegraphics[scale=0.29]{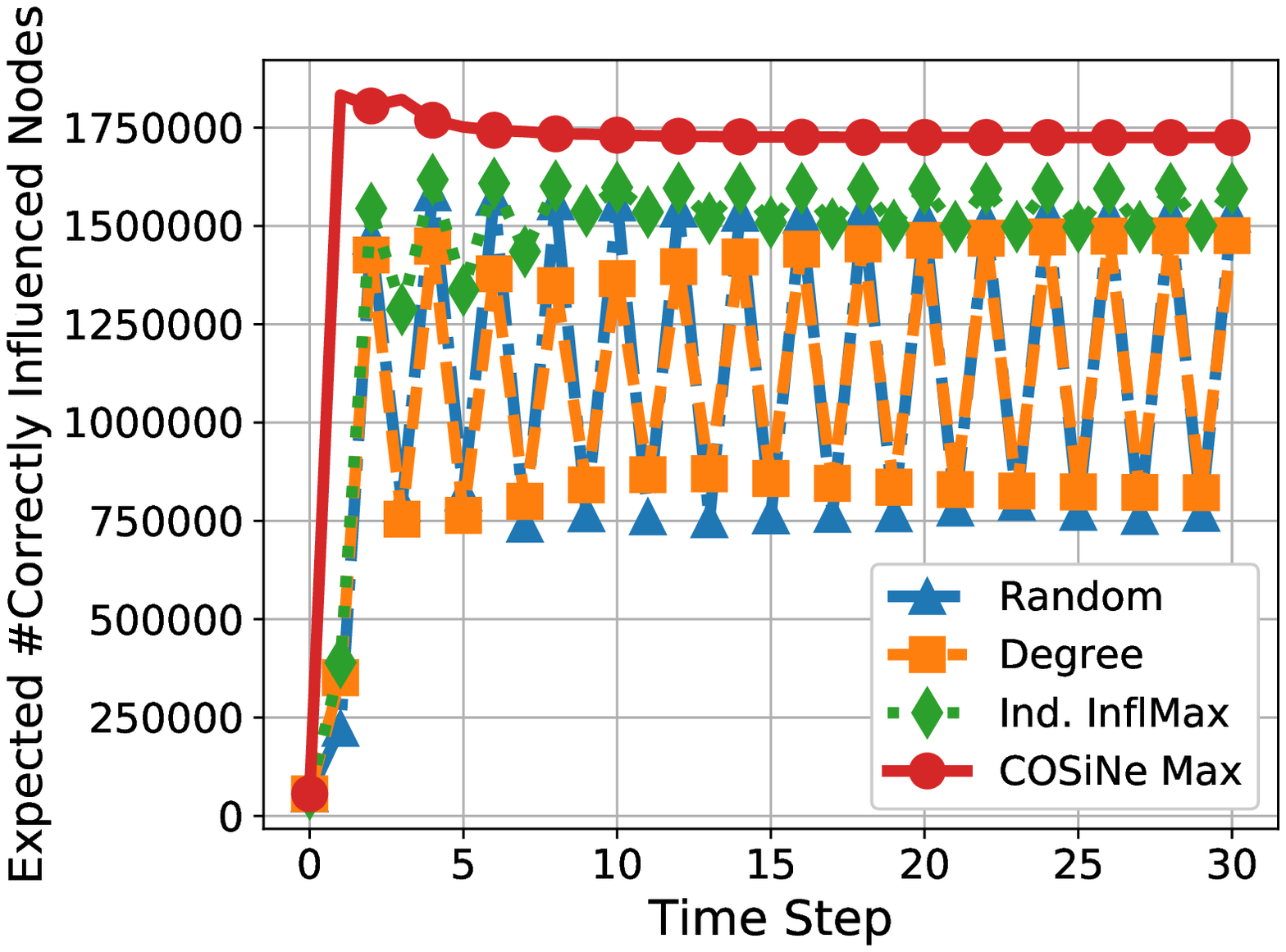}
\label{fig:boat_tg}
}
\vspace{-4.5mm}
\caption{\small Expected number of correctly influenced users for different time steps. Seeds are selected according to various algorithms.}
\label{fig:boat1}
\vspace{-4mm}
\end{figure*}
\begin{figure*}[t!]
\centering
\subfigure [{\small {\em Epinions}, \#seeds=5\% of all users}]{
\includegraphics[scale=0.29]{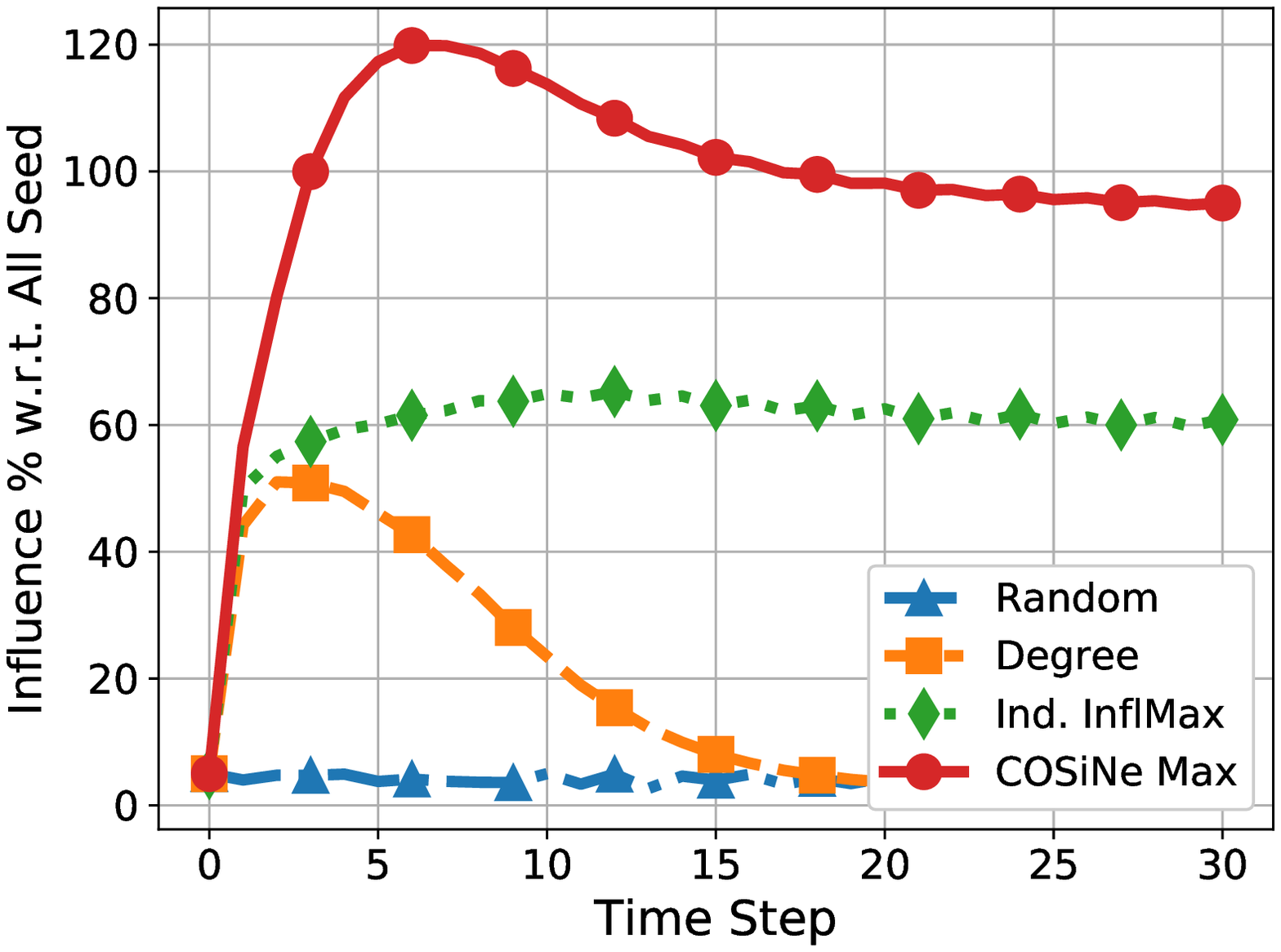}
\label{fig:boat2_ep}
}
\subfigure [{\small {\em GitHub}, \#seeds=5\% of all users}] {
\includegraphics[scale=0.29]{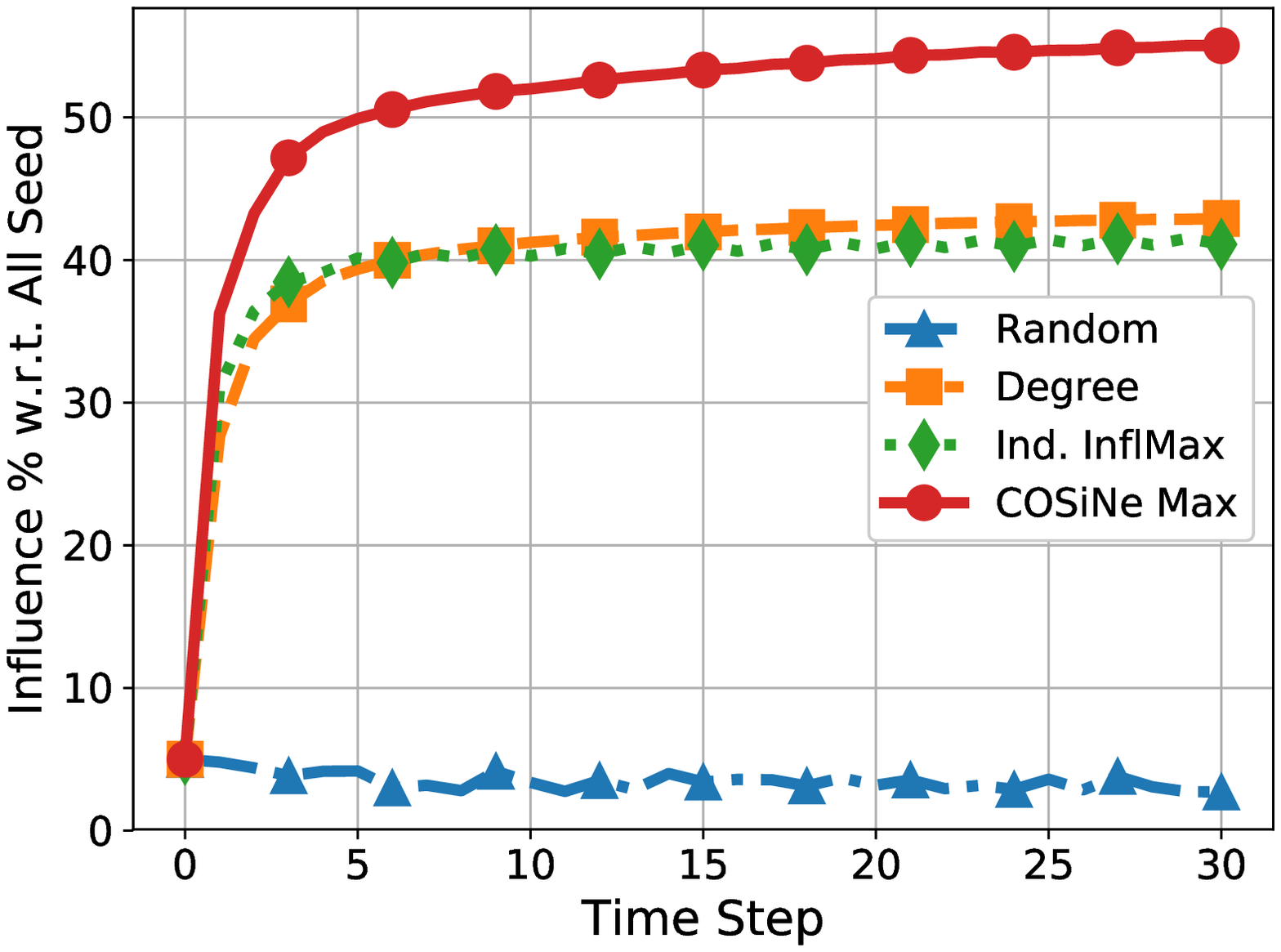}
\label{fig:boat2_gh}
}
\subfigure [{\small {\em Tagged}, \#seeds=1\% of all users}] {
\includegraphics[scale=0.29]{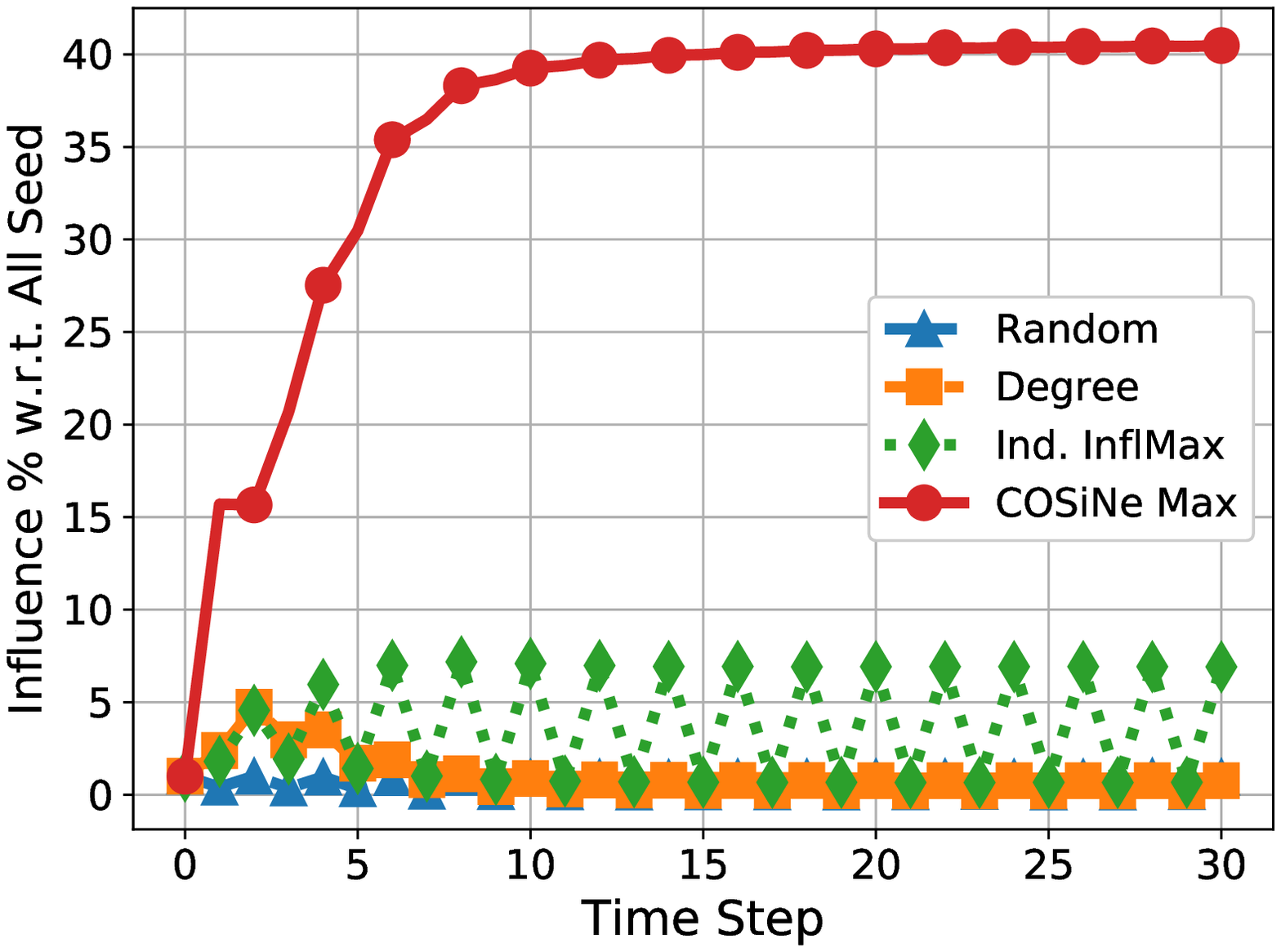}
\label{fig:boat2_tg}
}
\vspace{-4.5mm}
\caption{\small Influence percentage w.r.t. ``All Seed'' for different time steps. Seeds are selected according to various algorithms. ``All Seed'' denotes the case when all target nodes
are used as seeds, and influenced by the respective idea at $t=0$ (this metric is defined in Section~\ref{sec:metric}).}
\label{fig:boat2}
\vspace{-4mm}
\end{figure*}
\begin{figure*}[t!]
\centering
\subfigure [{\small {\em Epinions}, \#seeds=5\% of all users}]{
\includegraphics[scale=0.29]{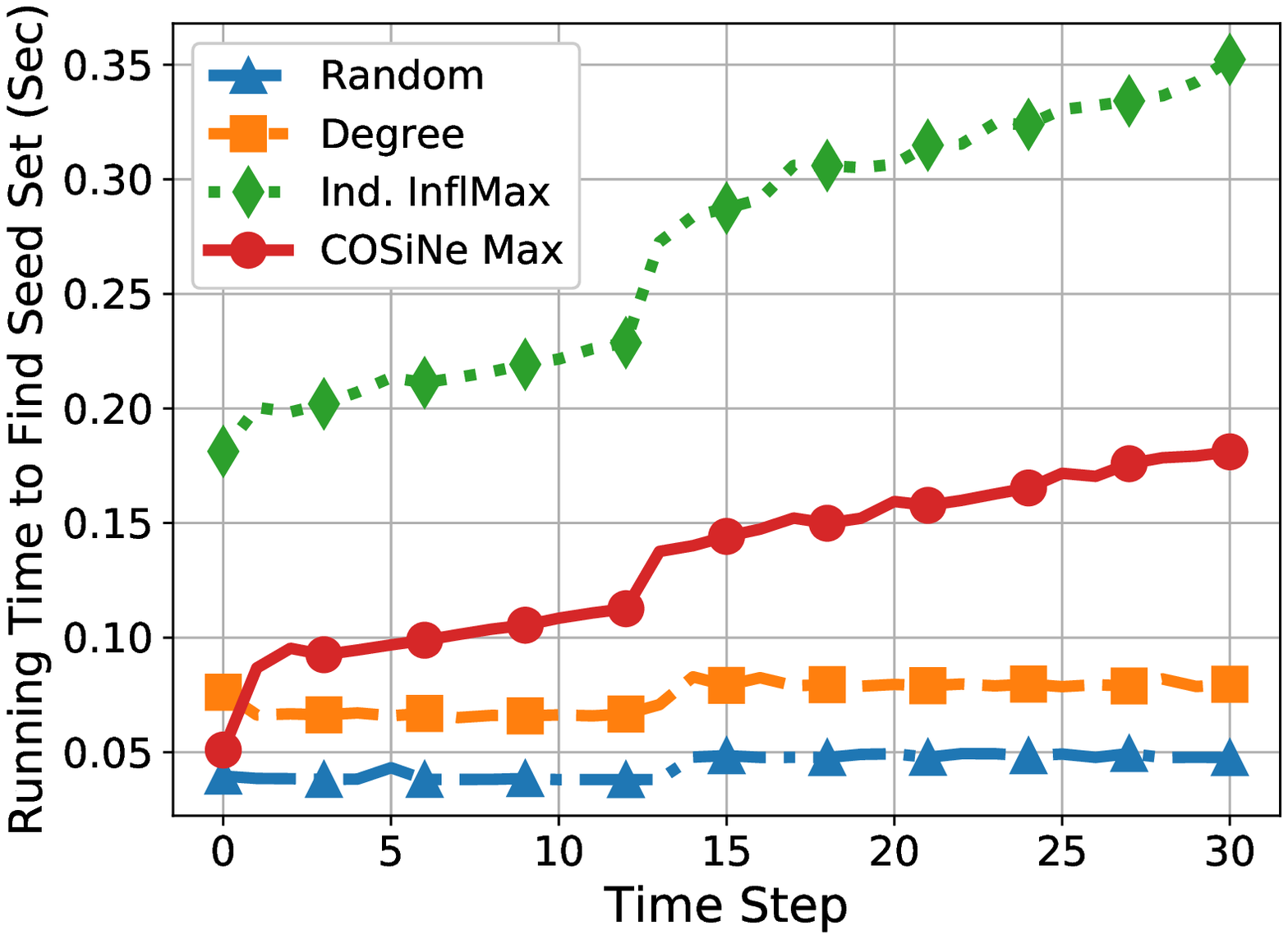}
\label{fig:boat3_ep}
}
\subfigure [{\small {\em GitHub}, \#seeds=5\% of all users}] {
\includegraphics[scale=0.29]{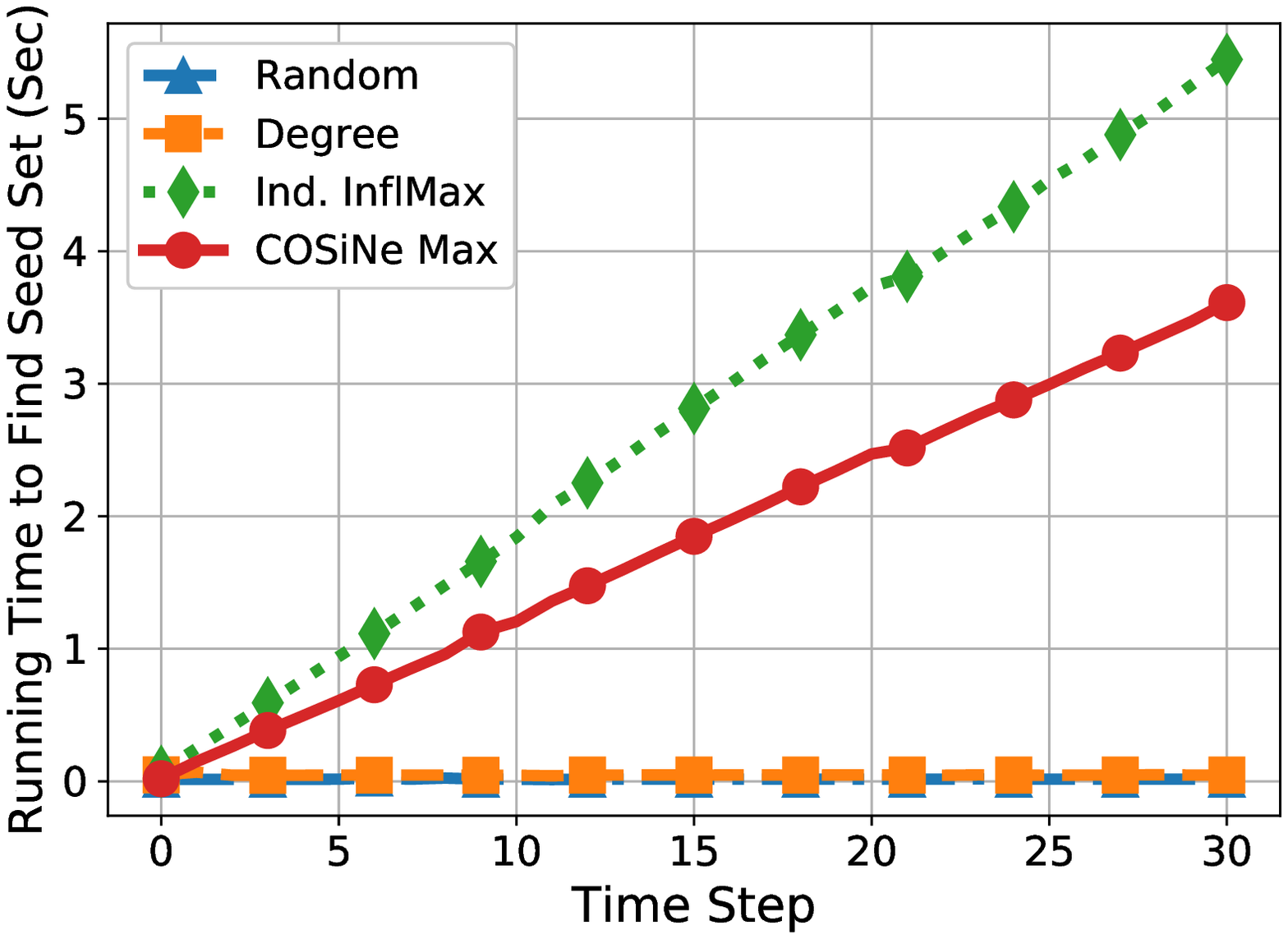}
\label{fig:boat3_gh}
}
\subfigure [{\small {\em Tagged}, \#seeds=1\% of all users}] {
\includegraphics[scale=0.29]{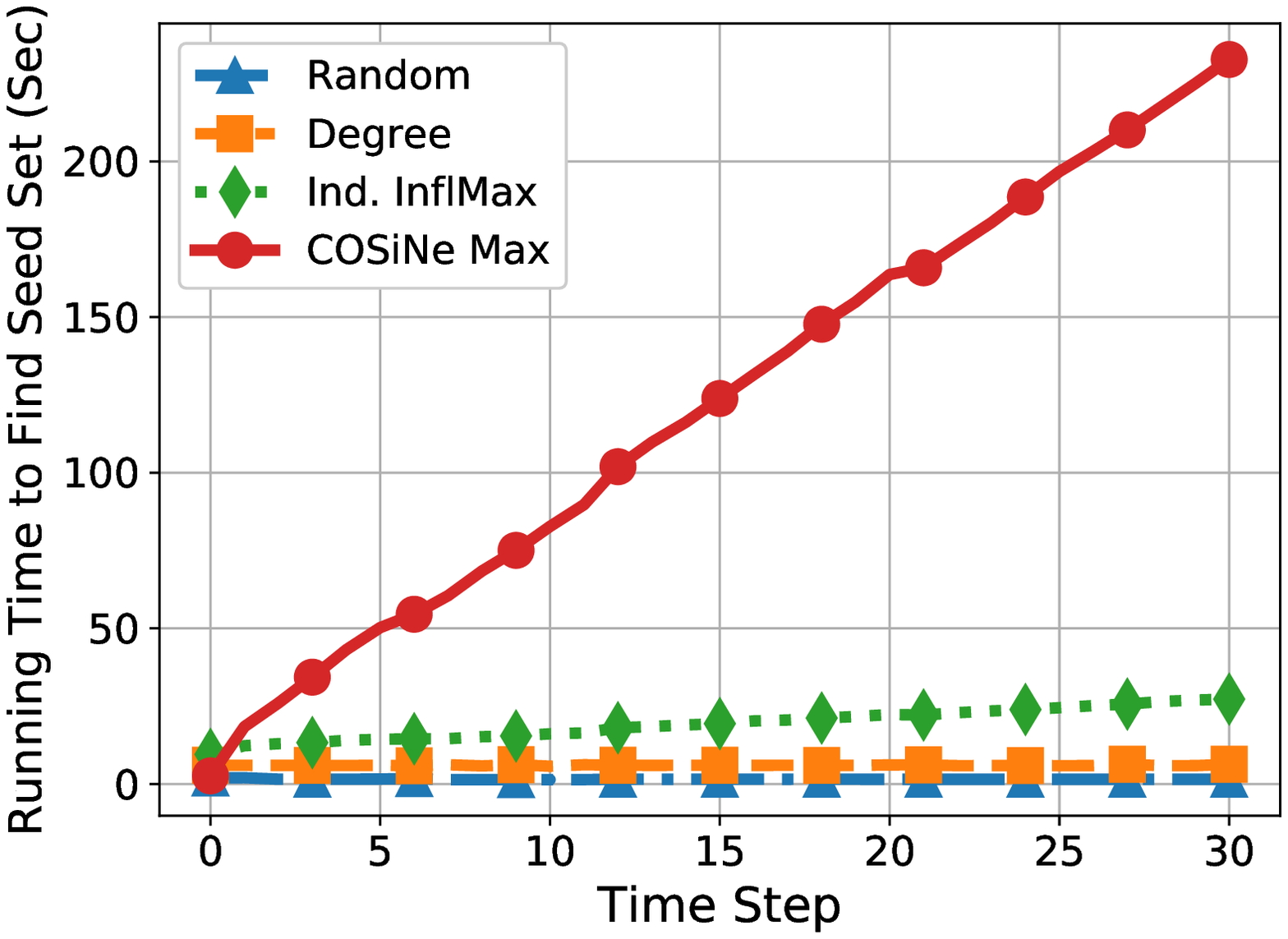}
\label{fig:boat3_tg}
}
\vspace{-4.5mm}
\caption{\small Running time to find seed nodes according to various algorithms.}
\label{fig:boat3}
\vspace{-4mm}
\end{figure*}
\vspace{-1.5mm}
\subsubsection{Parameters Setup}
\spara{\#Seeds.} We set the default number of seed nodes as 5\% for {\em Epinions} and {\em GitHub}, while 1\% for {\em Tagged}. This roughly translates to
7K, 1.3K, and 56K seeds in {\em Epinions}, {\em GitHub}, and {\em Tagged}, respectively. For sensitivity analysis, we vary the number of seeds from 0.8\% to 90\% (i.e., 1K to 120K) in {\em Epinions}.

\vspace{-1mm}
\spara{\#Target nodes.} In the experimental setting, we consider all nodes in the network as the target set of the campaigner. For sensitivity analysis, we
vary the number of target nodes from 15\% to 90\% (i.e., 20K to 120K) in the {\em Epinions} dataset. The target nodes are selected uniformly at random, and then we
split them into two non-overlapping partitions $V_1$ and $V_2$ based on the categories of products that each user reviews.

\vspace{-1.1mm}
\spara{Time steps.} We consider time steps up to 30 (short-term); for the long-term scenario we exhibit up to 500 time steps.
\vspace{-0.5mm}
\subsubsection{Evaluation Metrics}
\label{sec:metric}
We employ two metrics for the effectiveness measure.

\vspace{-1mm}
\spara{Expected number of correctly influenced nodes.}
We compute the number of nodes influenced by idea $O_1$ in target partition $V_1$, and by $O_2$ in target partition $V_2$.
Recall that  the probability of node $i$ adopting idea $O_1$ at time $t$ is defined as $p(O_1) = \frac{1+C_t(i)}{2}$,
and the probability of $i$ adopting idea $O_2$ at time $t$ is $p(O_2) = \frac{1-C_t(i)}{2}$.
Here, $C_t(i) \in [-1,1]$ is computed following Equation~\ref{eq:matrix_form}.

Moreover, we disregard weakly influenced nodes, i.e., node $i\in V_1$ when its $p(O_1)$ is less than a predefined threshold (0.5),
and $i\in V_2$ when its $p(O_2)$ is less than a predefined threshold (0.5). Such a user is likely to be undecided
between two opposite opinions on a specific issue. Formally, we report the following.
\vspace{-1.2mm}
\begin{equation}
\begin{aligned}
& \text{Expected number of correctly influenced nodes} \nonumber\\
\displaystyle &=\Sigma_{i \in V_1, C_t(i)>0} \left(\frac{1 +C_t(i)}{2}\right)+\Sigma_{i \in V_2, C_t(i)<0} \left(\frac{1-C_t(i)}{2}\right) \nonumber
\end{aligned}
\end{equation}
\begin{figure*}[tb!]
\centering
\subfigure [{\small Time step $t=3$}]{
\includegraphics[scale=0.25]{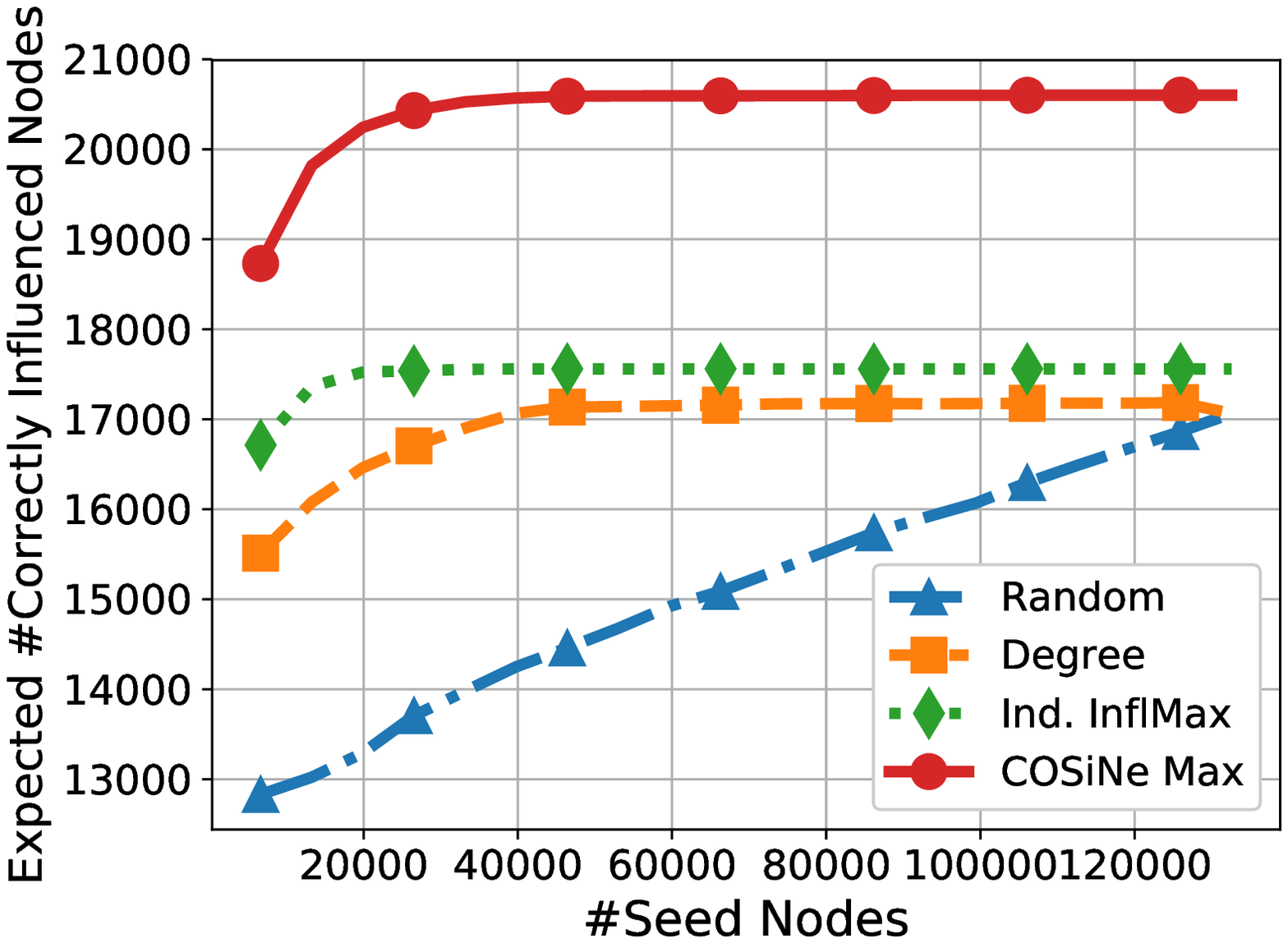}
\label{fig:sens_seed1}
}
\subfigure [{\small Time step $t=200$}] {
\includegraphics[scale=0.25]{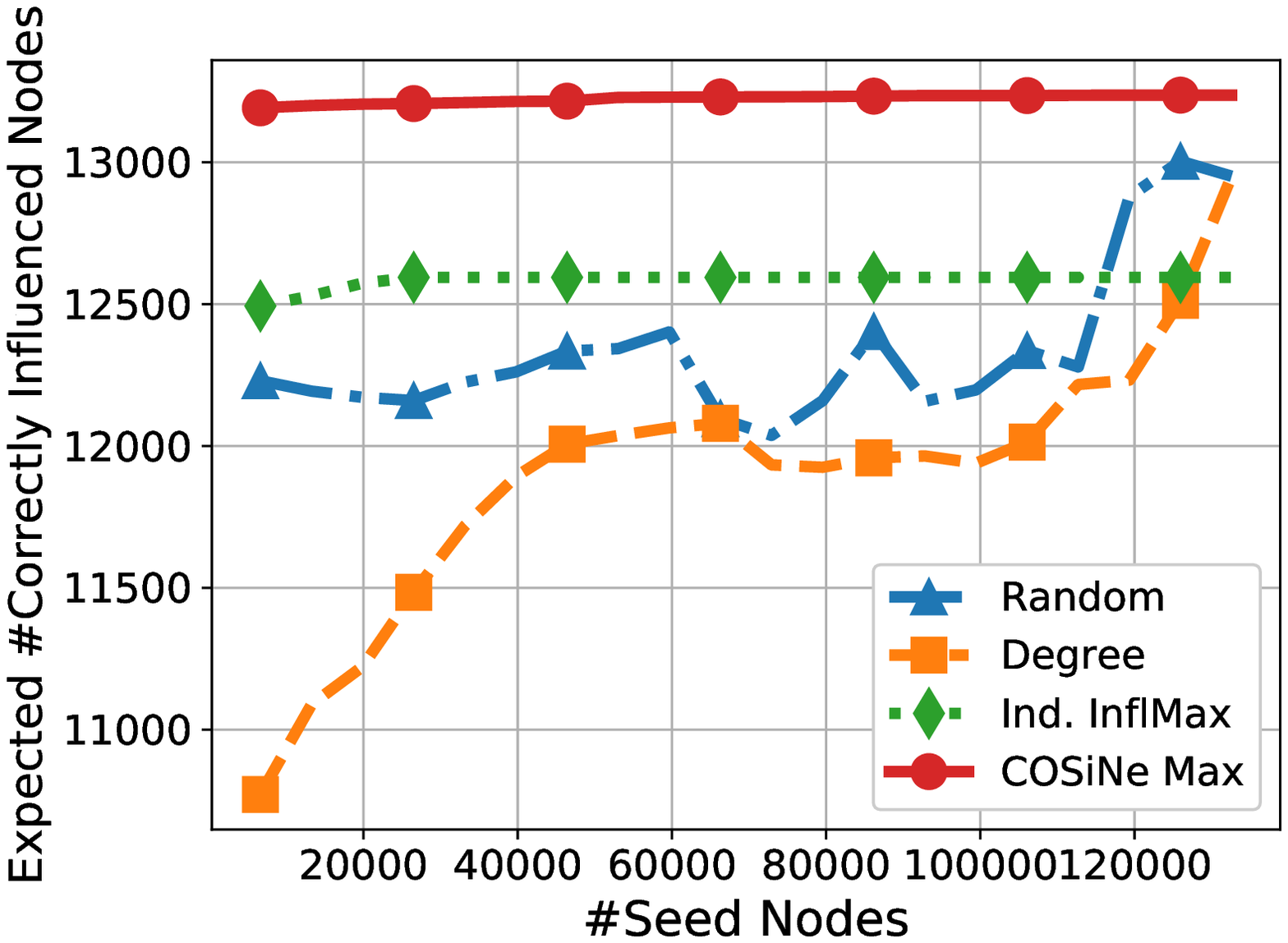}
\label{fig:sens_seed2}
}
\subfigure [{\small Time step $t=3$}] {
\includegraphics[scale=0.25]{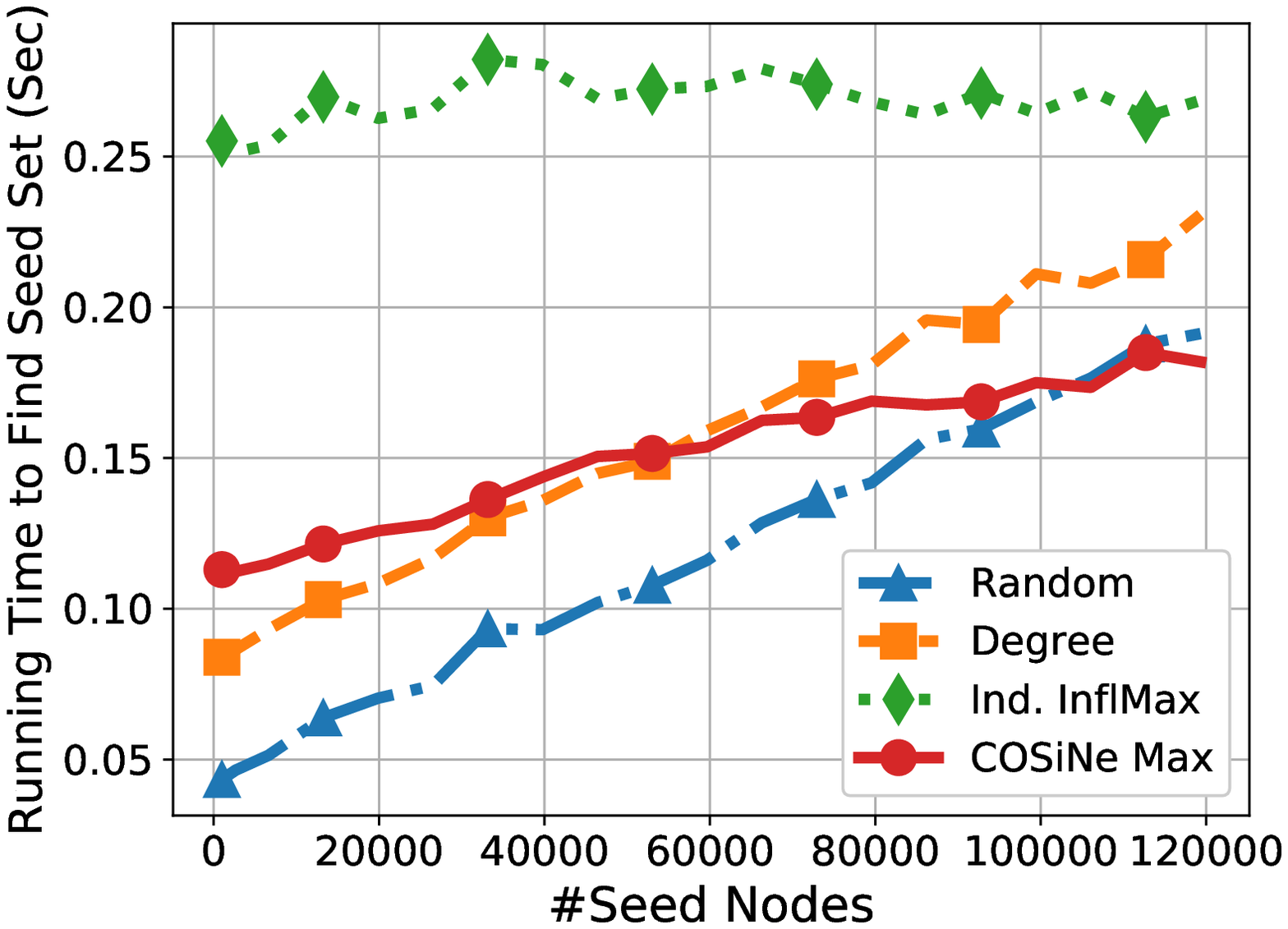}
\label{fig:sens_seed3}
}
\subfigure [{\small Time step $t=200$}] {
\includegraphics[scale=0.25]{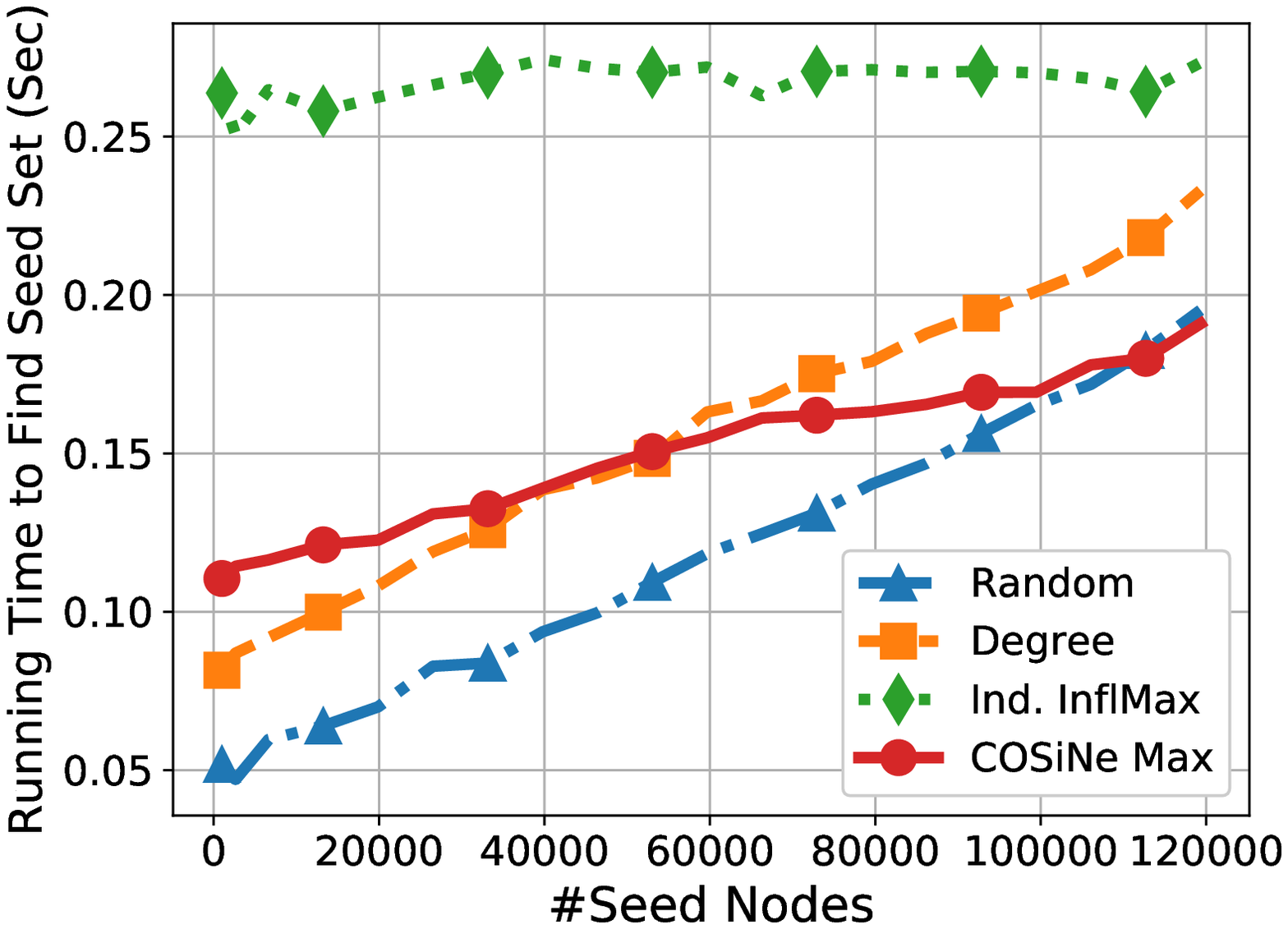}
\label{fig:sens_seed4}
}
\vspace{-4.5mm}
\caption{\small Sensitivity analysis w.r.t. varying number of seed nodes, {\em Epinions}.}
\label{fig:sens_seed}
\vspace{-4mm}
\end{figure*}
\begin{figure*}[tb!]
\centering
\subfigure [{\small Time step $t=3$}]{
\includegraphics[scale=0.25]{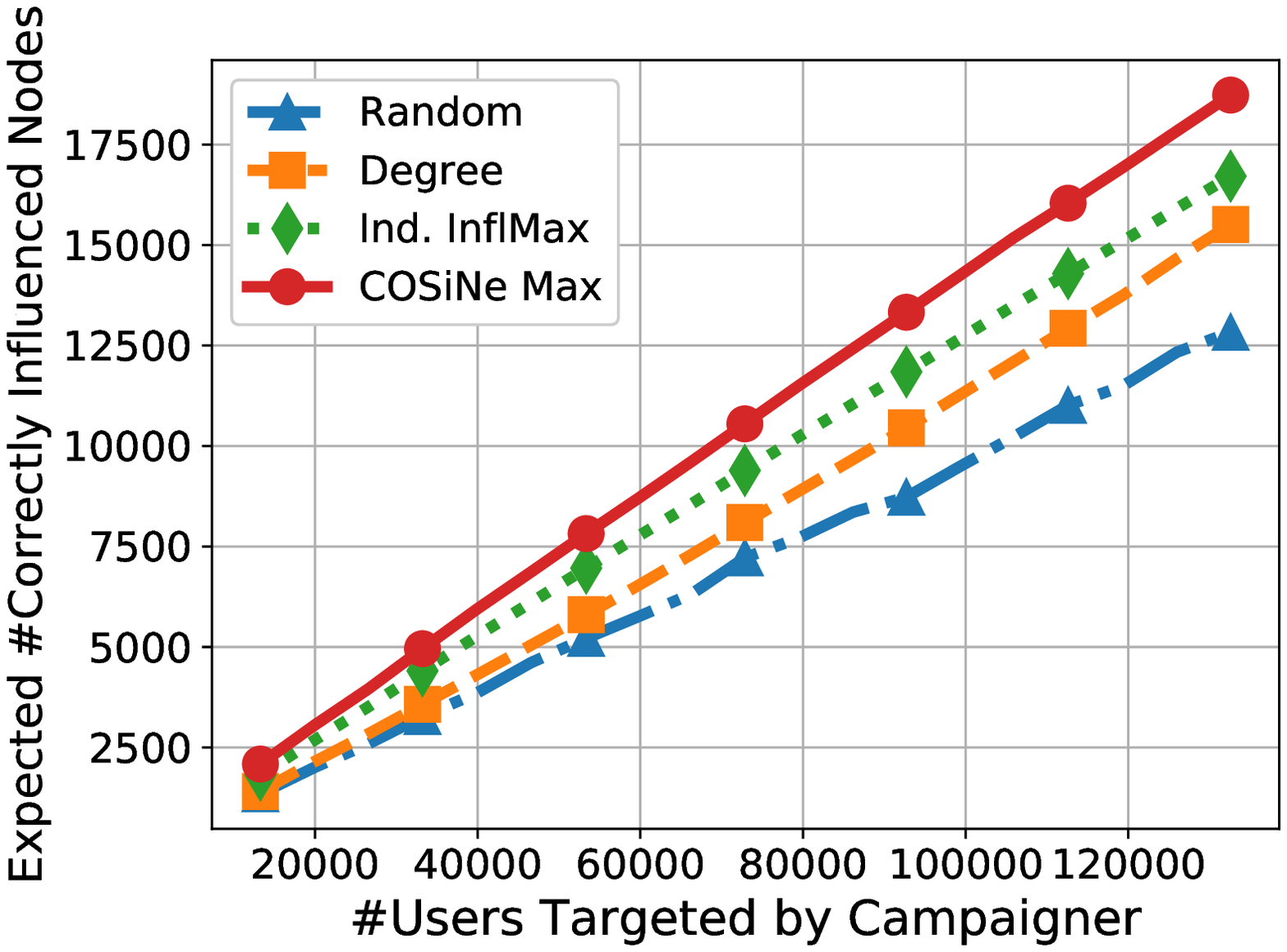}
\label{fig:sens_target1}
}
\subfigure [{\small Time step $t=200$}] {
\includegraphics[scale=0.25]{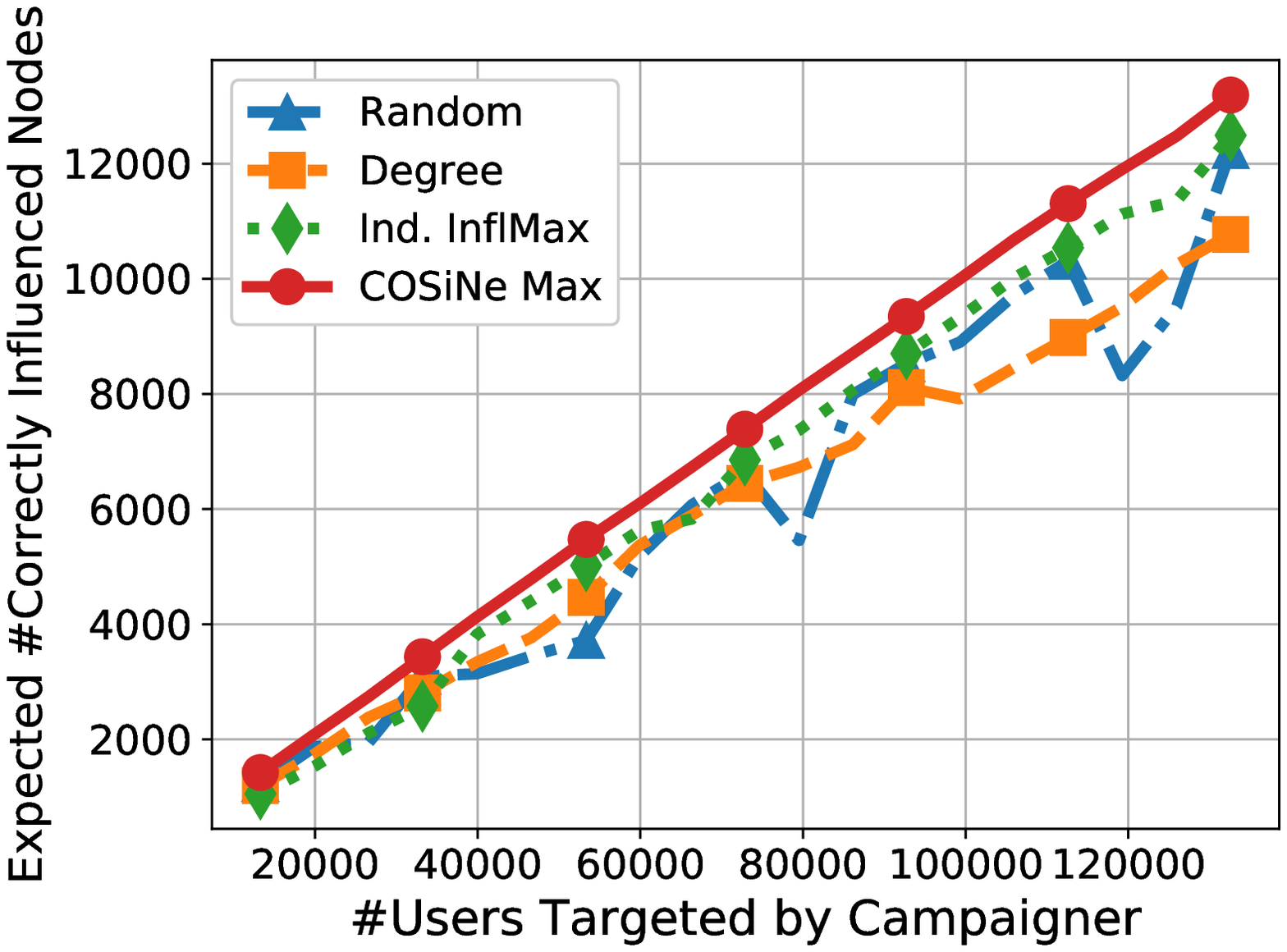}
\label{fig:sens_target2}
}
\subfigure [{\small Time step $t=3$}] {
\includegraphics[scale=0.25]{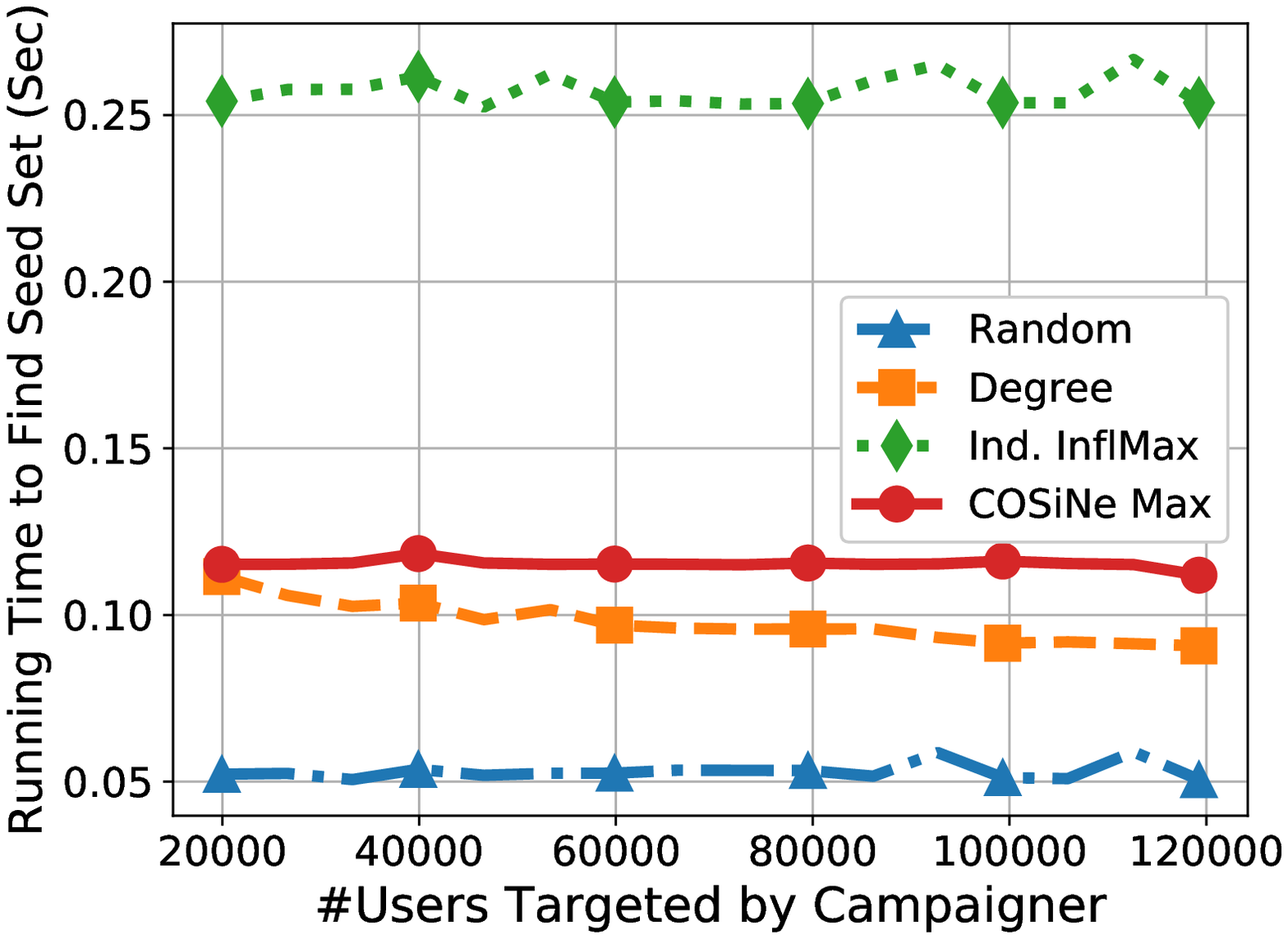}
\label{fig:sens_target3}
}
\subfigure [{\small Time step $t=200$}] {
\includegraphics[scale=0.25]{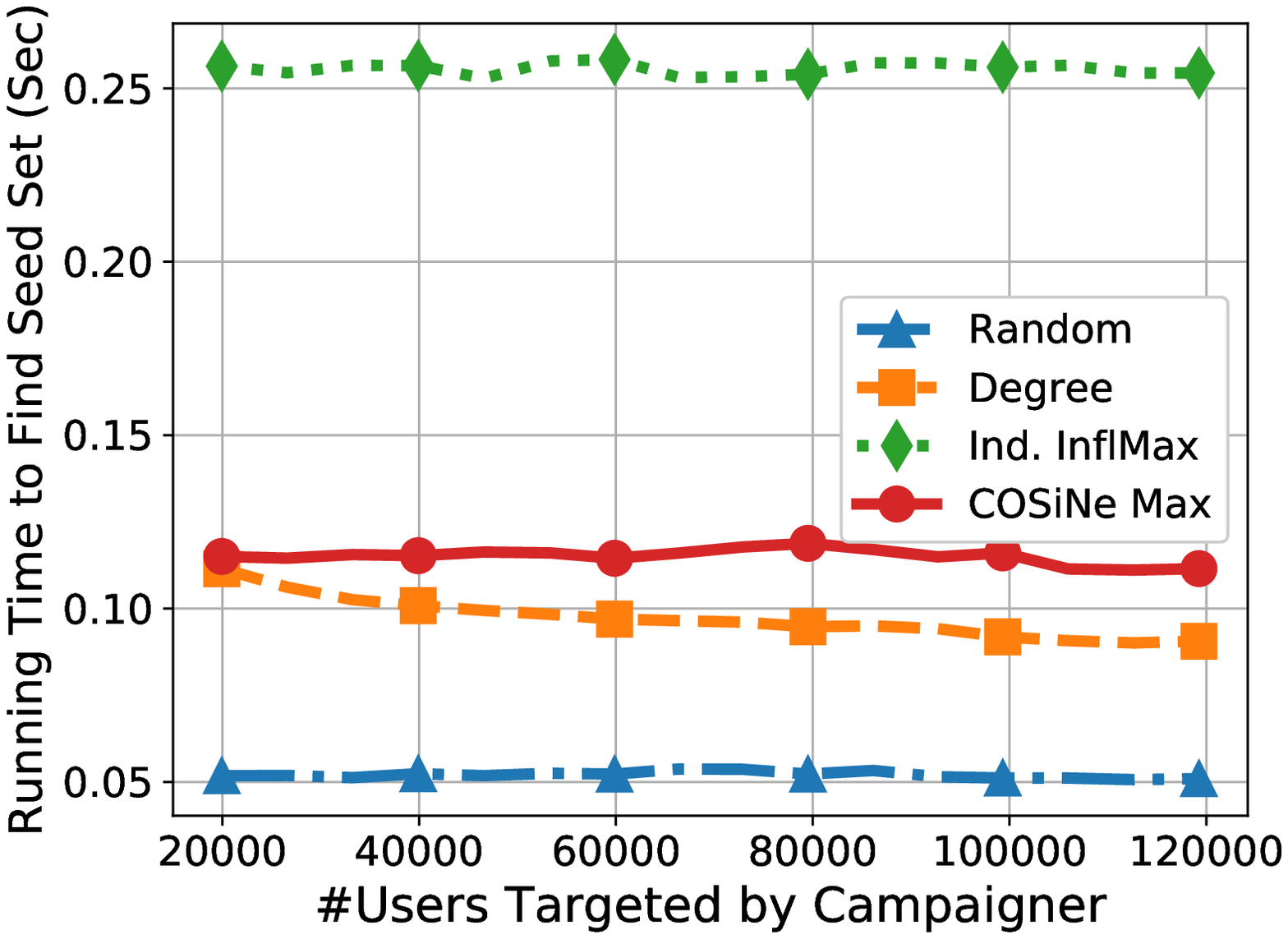}
\label{fig:sens_target4}
}
\vspace{-4.5mm}
\caption{\small Sensitivity analysis w.r.t. varying number of target nodes, {\em Epinions}.}
\label{fig:sens_target}
\vspace{-6mm}
\end{figure*}

\vspace{-2mm}
\spara{Influence percentage w.r.t. all targets as seeds.}
We also measure {\em campaign effectiveness constrained by a limited number of seeds}, with respect to the hypothetical scenario when all target nodes can be
employed as seeds. We recall that in Section~\ref{sec:problem_formulation}, the effectiveness of the campaign was formulated as ${\bm \rho^T} \cdot {\bf C_t}$.
This promotes opinion $O_1$ in partition $V_1$ and opinion $O_2$ in partition $V_2$, while penalising the reverse situation, that is, $O_1$ in $V_2$ and $O_2$ in $V_1$.

To better compare the aforementioned campaign effectiveness of each baseline and our proposed algorithm, we compare it to the case when all target nodes are assigned as seed nodes.
At time step $t=0$, the seeds are influenced with the respective idea of the target partition that they belong to. According to the voter model,
opposite influences on the same node cancel each other out, thus there could be a decay with time in the magnitude of influence.
Let us denote by $T_t$ the campaign effectiveness at time step $t$ in this scenario (i.e., when all target nodes were seed nodes at $t=0$).

Finally, we report $(\frac{{\bm \rho^T} \cdot {\bf C_t}}{T_t}\times 100)\%$
as the influence percentage w.r.t. all target nodes used as seed nodes.
\vspace{-2.5mm}
\subsection{Effectiveness Results}
\vspace{-1mm}
We present effectiveness results on three networks (Figure~\ref{fig:boat1}).
We find that our designed {\sf COSiNeMax} achieves higher expected number
of influenced nodes than all three baselines. Notice that {\em Epinions}
(Figure~\ref{fig:boat_ep}) shows some reduction in the expected number of correctly influenced nodes
with larger time steps till it saturates.  Such reduction is not observed in {\em GitHub} and {\em Tagged}.
This is due to higher sparsity of {\em Epinions}, with the presence of many separated components,
each consisting of a few nodes. In such a sparse network, random walks from seed nodes initially influence a large number of nodes.
However, this influence is unable to sustain at later time steps due to sparsity of the graph.
In other words, the sparsity of the network prevents long random walks from returning to the same nodes, thereby
reducing the influence over time.

When we compare the influence percentage (w.r.t. all targets as seeds) of each algorithm, {\sf COSiNeMax}
also outperforms all baselines (Figure~\ref{fig:boat2}). However, the peak value obtained in each dataset is different, with {\em Epinions} having the highest at 120\%,
{\em GitHub} having 55\%, and {\em Tagged} at 40\%.
The sparsity of {\em Epinions} dissipates the total influence $T_t$ very rapidly, reducing it by almost 75 \% in the first time step itself.
This quick decrease in influence is prevented with {\sf COSiNeMax} by selecting the seed nodes more intelligently,
thus achieving the peak value at higher than 100\%.

The oscillatory plots of the baselines in {\em Tagged} (Figures~\ref{fig:boat_tg}, \ref{fig:boat2_tg})
can be explained based on graph structure and node partitions. {\em Tagged} has more than three
times as many positive inter-partition edges than all other kinds of edges combined, thereby making these
partitions close to socially anti-balanced partitions. Thus, if the seed nodes in the two partitions are not
targeted by $O_1$ or $O_2$ intelligently, as it is done in case of baselines (see Section~\ref{sec:compete}),
such oscillatory behaviour in influence spread arises. This is similar to the oscillatory behaviour discussed in Section~\ref{sec:alg_long}
due to socially anti-balanced graph partitions. {\sf COSiNeMax}
is able to circumvent this problem by targeting all seed nodes in $V_1$ as $O_1$ when maximizing influence for even time steps, and as $O_2$ when maximizing influence for odd time
steps.
\vspace{-2.5mm}
\subsection{Efficiency Results}
\vspace{-1mm}
We compare running time to find seed nodes by all algorithms in Figure~\ref{fig:boat3}.
While time taken increases almost linearly with time steps for both {\sf COSiNeMax} and {\sf Individual InfMax},
it is evident that both {\sf Random} and {\sf Degree} are faster, and their seed set finding times are independent of
input time step.

In case of {\sf Individual InfMax}, the seed nodes are computed in two stages: once for opinion $O_1$ in the target set $V_1$,
and then for opinion $O_2$ in the target set $V_2$. However, {\sf COSiNeMax} holistically identifies all seed nodes in the entire graph.
This explains why {\sf COSiNeMax} is faster than {\sf Individual InfMax} over two smaller graphs. On the other hand, {\sf COSiNeMax}
requires more time than {\sf Individual InfMax} over {\em Tagged}, which is a larger dataset and the complexity of performing random walks
over entire graph dominates seed set finding time.
\begin{figure*}[tb!]
\centering
\subfigure [{\small Expected \#correctly inf. users}]{
\includegraphics[scale=0.29]{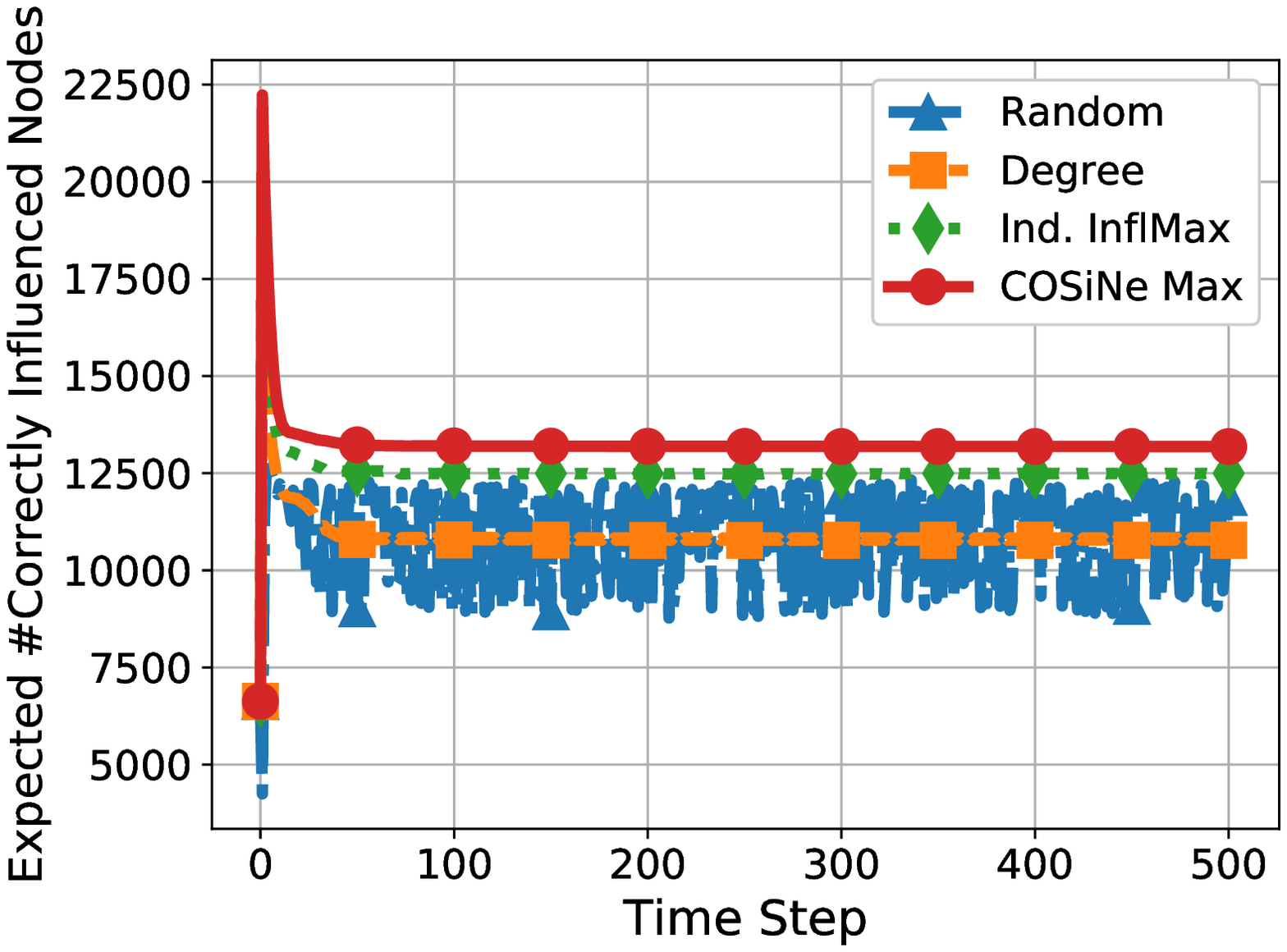}
\label{fig:boat4}
}
\subfigure [{\small Influence \% w.r.t. ``All Seed''}] {
\includegraphics[scale=0.29]{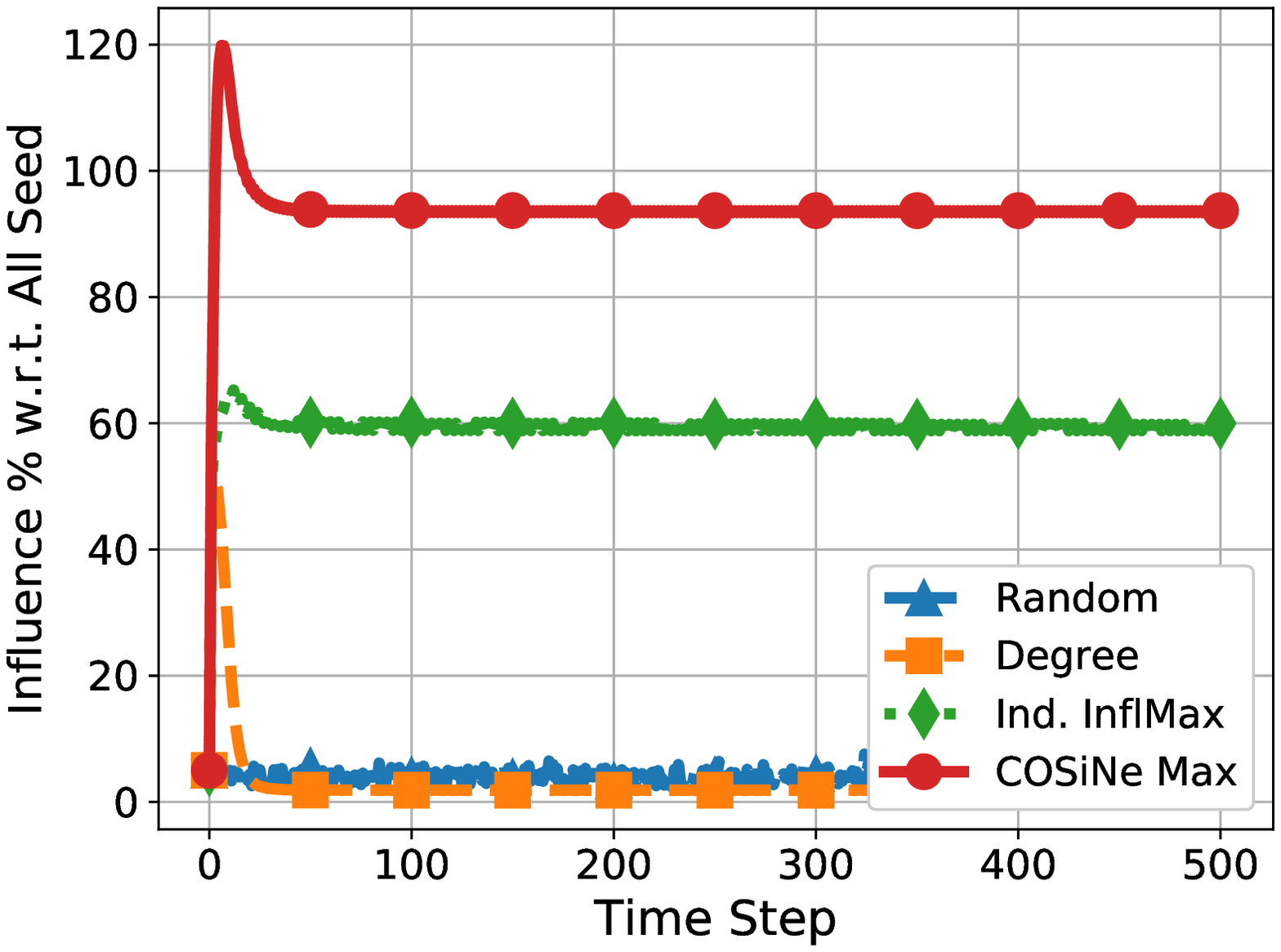}
\label{fig:boat5}
}
\subfigure [{\small Running time}] {
\includegraphics[scale=0.29]{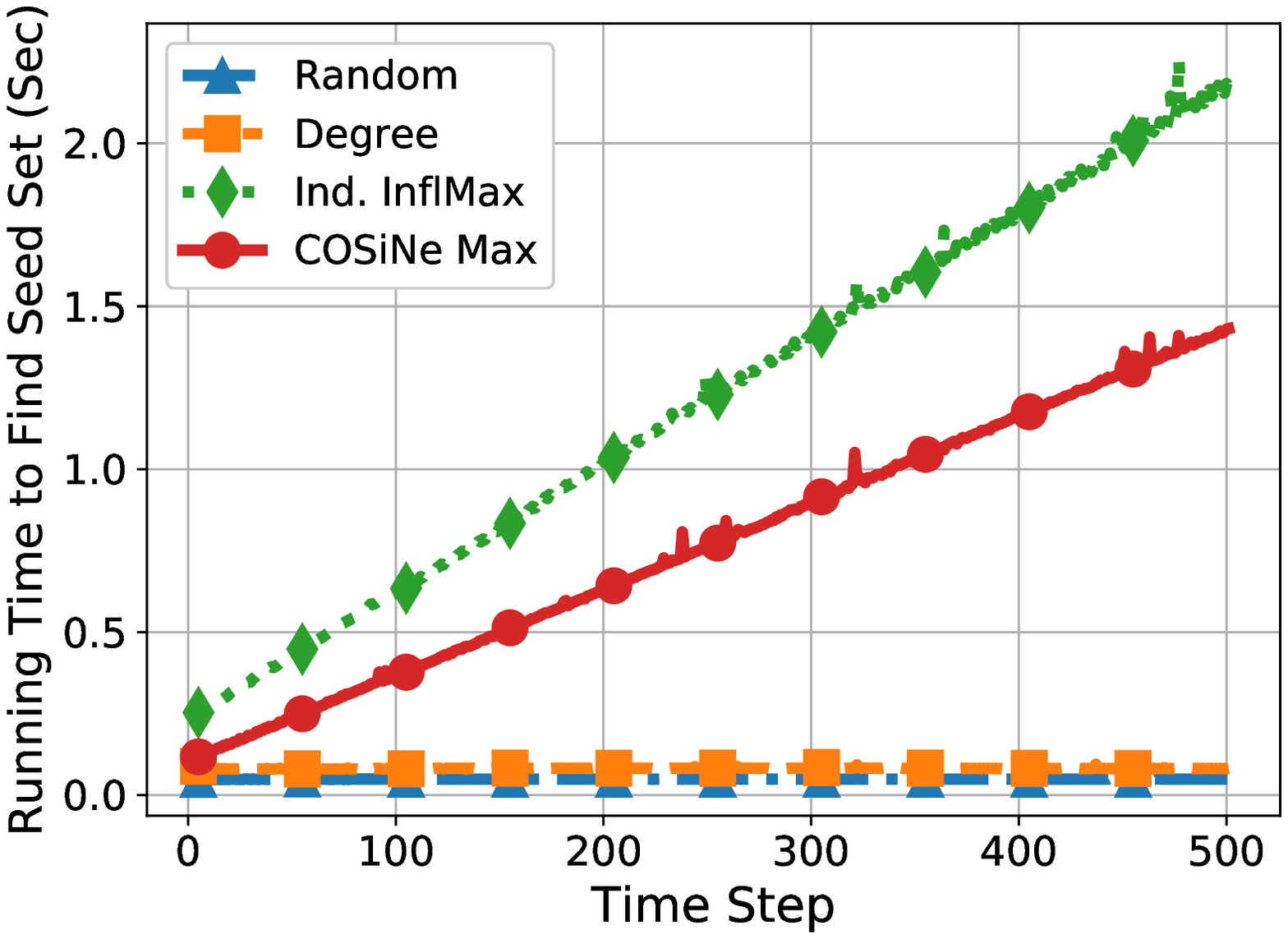}
\label{fig:boat6}
}
\vspace{-4.5mm}
\caption{\small Results on long-term opinions formation, {\em Epinions}, \#seeds=1\% of all users.}
\label{fig:boat_long}
\vspace{-6mm}
\end{figure*}
\vspace{-3mm}
\subsection{Sensitivity Analysis w.r.t. \#Seeds \& \#Targets}
\vspace{-1mm}
We investigate sensitivity of the algorithms w.r.t. numbers of seed and target nodes.
In Figures~\ref{fig:sens_seed} and \ref{fig:sens_target}, we present sensitivity analysis results using {\em Epinions},
generated for two time steps, $t=3$ (short-term) and $t=200$ (long-term). Finally, we also revisit the variation with time steps,
and study longer-term dynamics, with time steps from $0$ to $500$ (Figure~\ref{fig:boat_long}).

We find that the superior performance of our algorithm, {\sf COSiNeMax} --- both in terms of (a) expected number of correctly
influenced nodes and (b) influence percentage (w.r.t. all targets as seeds) --- is maintained for all parameter configurations.
Our empirical results demonstrate that {\sf COSiNeMax} finds the best quality solution regardless of the target set size, seed set budget, and
input time step.

In regards to long-term dynamics, we find that all algorithms, except the {\sf Random} baseline, achieves saturation over time, with no further variation in influence.
The expected number of correctly influenced nodes and the influence percentage (w.r.t. all targets as seeds) in this saturated state are both higher for our
 {\sf COSiNeMax} than the baselines.

\vspace{-2mm}
\section{Related Work}
\label{sec:related}

\vspace{-1.5mm}
\spara{Influence maximization in social networks.}
The classic influence maximization problem finds a limited number of seed users
that generate the largest expected influence cascade in a social network.
Kempe et. al. \cite{KKT03} designed the linear threshold (LT) and the independent cascade (IC) models,
and developed {\em approximation algorithms} having theoretical performance guarantees.
However, the computation of influence cascade is still \sharpP-hard following both IC and LT models \cite{CWW10}.
Lappas et. al. introduced the concept of target marketing and $k$-effectors ---
by identifying $k$ seed nodes such that a given activation pattern can be established \cite{LTGM10}.
%

\vspace{-1mm}
\spara{Competitive Influence maximization.}
Influence maximization in the presence of a negative campaign was investigated
in \cite{BKS07}, which assumes that the later campaign has prior knowledge
of rival side's initial seed nodes. Bordin et. al. \cite{BFO10} analyzed the similar problem
under the {\sf LT} model; while \cite{CCC11} attempts at preventing
the spread of an existing negative campaign in the network.
However, as competitive new products from rival companies are often launched around the same time,
\cite{LBGL13,KZK16} considered influence maximization in the presence of multiple competing campaigners,
who promote their products in a social network around the same time. Complementary influence maximization
was proposed in \cite{LCL15} for promoting complementary products together.

Our work is fundamentally different from prior literature.
{\bf First}, they generally consider activation based models (e.g., IC and LT) suitable for {\em one-time} product purchase.
In contrast, our voter
model allows users to switch opinions at later times based on their neighbors' opinions.
Thus, voter model is more suitable to study opinion diffusion and formation in online social networks.
{\bf Second}, although earlier works consider multiple competitive campaigns, different from our study they do not consider
diffusion with both positive and negative edges in a {\em signed} social network. {\bf Third}, due to the inherent complexity
of IC, LT models and their variants, the problems investigated in those works are generally \NP-hard and also \sharpP-hard, while
the voter model can solve our problem {\em exactly} in linear time.

\vspace{-1mm}
\spara{Signed social networks.}
Signed network research dates back to 1940's with the work of Heider \cite{H46}, and was
formalized by Harary and Carwright \cite{cartwright56}. Signed networks have recently
become popular in data mining and social network analysis (for a survey, see \cite{Tang16}).
In \cite{Leskovec10}, Leskovec et al. studied the structure of social networks with negative relationships
based on two social science theories --- balance theory and status theory.
Kunegis et al.\cite{KSLLLA10} investigated spectral properties of signed undirected networks,
having applications in link predictions and clustering. Tang et al. \cite{Tang16} performed
node classification in signed networks.

\vspace{-1mm}
\spara{Influence maximization in signed social networks.}
With the prevalence of signed social networks, recent works investigated the problem of finding the seed set that maximizes
positive influence, which is also known as positive influence maximization. \cite{LXCGSL14,Shen15,Srivastava15}
studied positive influence maximization under different extensions of IC and LT models. Li et al. \cite{VoterModelSigned} explored similar problem
in a signed social network with voter model. Unlike ours, they do not aim at maximizing two contrasting opinions
in two non-overlapping target regions. Moreover, in \cite{VoterModelSigned} all seed nodes can be influenced by only one type of idea, that is,
for positive influence maximization, all seeds will be influenced by the positive idea.
However, as demonstrated in our experiments, maximizing each influence {\em separately} (i.e., Individual InfMax) results in a sub-optimal
solution compared to ours (i.e., {\sf COSiNeMax}): We return optimal seed nodes considering the spread of two contrasting ideas {\em simultaneously}.

%

\vspace{-1mm}
\spara{Measuring and minimizing social polarization.} 
Garimella et al. detected topics from Twitter data that caused intense debate \cite{GarimellaW17}.
Techniques to reduce polarization and disagreement in social networks by updating nodes and edges were developed in \cite{GMGM17,mossel2017,MuscoMT18}.
We acknowledge that in certain situations it is indeed necessary to reduce polarization, as otherwise created ``echo chambers'' (a metaphoric situation
in which specific kinds of opinions and convictions are strengthened and spread through the repetition and continuous communication among users who share the same
kind of thoughts inside a {\em closed} system) may result in extreme conflicts and instability. However, as we discussed earlier, for
public awareness, open and honest discussion, diversity and inclusion, educated voting, and towards better democracy,
{\em polarization, with certain regulations, is the key} \cite{C19,S50,NatureWiki,Fortune,GovOpp, Columbia,Queens}. Our work is motivated from this perspective.

\vspace{-2mm}
\section{Conclusions}
\label{sec:conclusions}
We formulated and investigated the novel problem of contrasting opinions maximization
in two distinct target groups, respectively, over a signed social network. Motivated by scenarios such as increasing
voter engagement and turnout, steering public debates and discussions on societal issues with contentious opinions,
we adapted the voter model to effectively study influence diffusion. We efficiently solved
this problem, and designed an exact algorithm. We then empirically compared this algorithm with several baselines on three
real-world signed network datasets. Our analysis reveals that the proposed algorithm, {\sf COSiNeMax} finds the seed set
with the highest expected number of influenced nodes, and has the highest relative total influence.
This behaviour is demonstrated over all datasets and for different variations of time steps, seed set budget,
and target population size parameters. In future, it would be interesting to consider adaptive seeding,
as opposed to one-time seeding, for even more effective short-term opinions maximization in a signed, social network.



\end{document}